\documentclass[iop,revtex4,numberedappendix,openany]{emulateapj} %
\usepackage{float}
\usepackage{rotating}
\usepackage{graphicx}
\usepackage{chngpage}
\usepackage{amsmath}

\usepackage[caption=false]{subfig}

\shorttitle{Implications for $H_0$}
\shortauthors{Wilson et al.}

\begin{document}

\title{A Spectroscopic Survey of the Fields of 28 Strong Gravitational Lenses: Implications for $H_0$}

\author{Michelle L. Wilson\altaffilmark{1}, Ann I. Zabludoff\altaffilmark{1}, Charles R. Keeton\altaffilmark{2}, Kenneth C. Wong\altaffilmark{3,4,5}, Kurtis A. Williams\altaffilmark{6}, K. Decker French\altaffilmark{1}, and Ivelina G. Momcheva\altaffilmark{7}}

\altaffiltext{1}{Steward Observatory, University of Arizona, 933 North Cherry Avenue, Tucson, AZ 85721, USA}
\altaffiltext{2}{Department of Physics and Astronomy, Rutgers University, 136 Frelinghuysen Road, Piscataway, NJ 08854, USA}
\altaffiltext{3}{National Astronomical Observatory of Japan, 2-21-1 Osawa, Mitaka, Tokyo 181-8588, Japan}
\altaffiltext{4}{Institute of Astronomy and Astrophysics, Academia Sinica (ASIAA), P.O. Box 23-141, Taipei 10617, Taiwan}
\altaffiltext{5}{EACOA Fellow}
\altaffiltext{7}{Space Telescope Science Institute, 3700 San Martin Drive, Baltimore, MD 21218, USA} 
\altaffiltext{6}{Department of Physics and Astronomy, Texas A\&M University-Commerce, Commerce, TX, 75428, USA}

\begin{abstract}

Strong gravitational lensing provides an independent measurement of the Hubble parameter ($H_0$).  One remaining systematic is a bias from the additional mass due to a galaxy group at the lens redshift or along the sightline. We quantify this bias for more than 20 strong lenses that have well-sampled sightline mass distributions, focusing on the convergence $\kappa$ and shear $\gamma$. In 23\% of these fields, a lens group contributes a $\ge$1\% convergence bias; in 57\%, there is a similarly significant line-of-sight group.  For the nine time delay lens systems, $H_0$ is overestimated by 11$^{+3}_{-2}$\% on average when groups are ignored. In 67\% of fields with total $\kappa \ge$ 0.01, line-of-sight groups contribute $\gtrsim 2\times$ more convergence than do lens groups, indicating that the lens group is not the only important mass. Lens environment affects the ratio of four (quad) to two (double) image systems; all seven quads have lens groups while only three of 10 doubles do, and the highest convergences due to lens groups are in quads. We calibrate the $\gamma$-$\kappa$ relation:  $\log(\kappa_{\rm{tot}}) = (1.94 \pm 0.34) \log(\gamma_{\rm{tot}}) + (1.31 \pm 0.49)$ with a rms scatter of 0.34 dex. Shear, which, unlike convergence, can be measured directly from lensed images, can be a poor predictor of $\kappa$; for 19\% of our fields, $\kappa$ is $\gtrsim 2\gamma$. Thus, accurate cosmology using strong gravitational lenses requires precise measurement and correction for all significant structures in each lens field.

\end{abstract}

\keywords{galaxies: groups: general -- gravitational lensing: strong}

\section{Introduction}

Strong gravitational lensing has long been used to measure cosmological parameters, such as the Hubble constant through measuring time delays for systems with quasar sources \citep{refsdal64,schechter97,keeton97H0,koopmans03,saha06,oguri07,suyu10,suyu13,rathna15,birrer16,chen16,bonvin17,wong17} and the dark energy density through determining statistical properties of strong lensing systems \citep{turner90,chae03,mitchell05,oguri08}.  Lensed type Ia supernovae likewise can be used to measure $H_0$ \citep{refsdal64,goldstein17}, although few have been found yet \citep{kelly15,goobar17}.  Future large surveys, such as those using the Large Synoptic Survey Telescope (LSST) and \textit{Euclid}, will discover a large sample (i.e., a few thousand) of lensed quasars that could then be used to statistically constrain these parameters to high precision \citep{oguri10,laureijs11,linder11,lsstcolab12}.

Much of the work on galaxy-scale strong gravitational lenses has indicated that often the lens galaxy is not the only mass that significantly affects the lensing potential \citep[e.g.,][]{wallington96,keeton97,claeskens06,collett13}.  Since any mass along the line-of-sight (hereafter LOS) between the observer and the source can contribute to the lensing, any structure in which the lens galaxy might reside, as well as any unrelated structure sufficiently massive and close either in redshift or projected on the sky, might be important.  In addition, the lens environment might affect the lensed image morphology \citep[i.e., how many images of the source are present;][]{keeton04,momcheva06}.

Previous work has indicated that many galaxy-scale lenses are located in groups or clusters and/or in fields with projected LOS structures \citep[e.g.,][]{kundic97,fassnacht02,momcheva06,auger08,sluse16}.  However, only small footprints on the sky around a few galaxy-scale lens systems have been observed with deep follow-up spectroscopy.  So, how common these structures are is not well constrained.  If gravitational lensing is going to be one of the several methods used to determine $H_0$ to $<$ 1\% accuracy, which is also needed for upcoming dark energy studies \citep[e.g.,][]{linder11}, systematics such as those due to local environment and LOS mass must be quantified.

We use a spectroscopically-sampled subset of 26 of 28 galaxy-scale strong gravitational lenses \citep[][]{momcheva15,wilson16} to constrain the lens environments and LOS structures. We use the convergences and shears to assess the importance of groups.  The uncertainty due to lens environment and LOS structures is a systematic, rather than statistical, uncertainty.  Convergence, which affects the measurements of $H_0$, is not measured on the sky like shear, nor can it be inferred through models.  So, it is important to constrain uncertainties due to groups specifically.

In our analyses, we consider convergences or shears of individual groups or total lines-of-sight to be significant if they are $\ge$ 0.01.  We adopt this threshold because a convergence of 0.01 will correspond to a 1\% bias in $H_0$ \citep[as $h \propto (1 - \kappa)$,][]{keeton04}, the level of precision necessary for this method to independently constrain $H_0$ at the current leading levels of precision using other methods \citep[see, e.g.,][]{keeton04}.  Shears of 0.01 lead to $\sim 3\sigma$ effects, because shear enters the lens equation as $\gamma$ multiplied by the lensed image position, which usually has uncertainties $\sim 0.^{\prime\prime}$003 \citep[][]{courbin97}.  The uncertainty due to time delay measurements \citep[e.g.][]{suyu13} and constraints on the external convergence derived statistically from galaxy counts and cosmological simulations \citep[e.g.][]{rusu17} have been at the $\sim$ 1-2\% level.  In addition, we consider convergences and shears $\ge$ 0.05.  The 5\% level is about the level of discrepancy persisting between different methods of measuring $H_0$.

In this paper, we describe the data in Section \ref{sec:data} and our lensing formalism in Section \ref{sec:formalism}.  We then discuss the importance of groups to lensing in the order in which complexity historically has been added to lensing models when considering convergence, which is not measured on the sky: we first examine the lens environment (Section \ref{sec:lensenvirons}) then LOS structures (Section \ref{sec:losstructureeffects}).  We structure our discussion of shear in the same way for clarity; however, we note that total shear is measured on the sky, and historically increasing complexity has involved dividing that shear into a larger number of possible contributors (i.e., lens environment and LOS groups) rather than adding additional terms.  Additionally, we look for differences in environment and LOS structures for systems with quad and double image morphologies (Sections \ref{ssec:quadvsdoublesenv} and \ref{ssec:quadvsdoubleslos}, respectively).  We consider implications of the convergences due to groups for measuring $H_0$ in Section \ref{sec:H0}.  Finally, in Section \ref{sec:gammatotvkappatot}, we investigate the relationship between $\gamma_{\rm{tot}}$ and $\kappa_{\rm{tot}}$.

Throughout this paper, we adopt the values of $H_0$ = 71 km~s$^{-1}$ Mpc$^{-1}$, $\Omega_m$ = 0.274, and $\Omega_{\Lambda}$ = 0.726 \citep{hinshaw09}.

\section{The Data}
\label{sec:data}

\subsection{Sample Selection}

The lens sample selection is described in \citet{momcheva15}.  It depended in part on the ancillary data available at the time of spectroscopic observations as well as the lens fields' accessibility from the telescope site. Some early preference was given to fields with known large shears, ill-fitting models, and/or poorly characterized lens environments.

\citet{williams06} details the collection and reduction of the photometry in the 28 lens fields.  Galaxy redshifts were obtained by \citet{momcheva15} for a subset of the photometric objects in those fields; the​ spectroscopically well-sampled regions have radii ranging from $\sim 5^{\prime}$ to $\sim 20^{\prime}$ from the lens. These large sampled regions allow for the identification and accurate determination of group properties (i.e., group centroid and group velocity dispersion), which can significantly impact the lensing potential even when the group is projected a few arcminutes from the lens.

For 26 of the 28 fields with the most complete spectroscopic data, \citet{wilson16} create a group catalog including any groups projected along the sightline to the lens as well as groups in which the lens galaxy is itself a member. The properties of these 26 fields are in Table \ref{table:lenssystemprops}.  Here we use Wilson et al.'s group velocity dispersions, mean redshifts, and projected spatial centroids.  We focus primarily on the subset of 21 fields with lenses with firm spectroscopic redshifts.  We consider one additional lens field, b0712, which has both a group at the lens and a large foreground supergroup \citep{wilson16}, when analyzing lens environments (Sections \ref{sec:lensenvirons} and \ref{sec:H0}; Figure \ref{fig:kappaeffquaddoub}), but exclude it elsewhere due to the uncertain mass of the supergroup (see Section \ref{sec:formalism}).

\subsection{Photometry}

The Mosaic-1 imager on the Kitt Peak National Observatory (KPNO) Mayall 4 m telescope and the Mosaic II imager on the Cerro Tololo Inter-American Observatory (CTIO) Blanco 4 m telescope were used to collect images through the ``nearly Mould'' $I$-band filter and either the Harris $V$ filter or the Harris $R$ filter (depending on the redshift of the lens galaxy). SExtractor version 2.3.2's MAG\_AUTO \citep{bertin96} was used to determine the \citet{kron80} Vega magnitudes in the $I$-band.  Colors were measured by degrading one image to the resolution of the other then calculating the aperture magnitudes.    A calibration was applied to put the photometry on the Kron-Cousins filter system.  The photometry was not corrected to total magnitudes.  The star-galaxy separation limit is 21.5 mag.

\subsection{Spectroscopy}

The spectra were obtained using Hectospec on the MMT 6.5 m telescope and LDSS-2, LDSS-3, and IMACS on the Magellan 6.5 m telescopes.  A method based on the routine of \citet{cool08}, which fits measured spectra to templates and selects that with the lowest $\chi^2$, was used to measure the redshifts.  Objects in the photometry that were not followed up but that had redshifts in NED\footnote[1]{The NASA/IPAC Extragalactic Database (NED) is operated by the Jet Propulsion Laboratory, California Institute of Technology, under contract with the National Aeronautics and Space Administration.} were added to the redshift catalog.  The resulting catalog includes 10002 unique galaxy redshifts, with  79.4\% between $z=$ 0.1 and 0.7 and a median redshift of 0.360. %page13 of Iva's paper

\subsection{Group Catalog}

\subsubsection{Group Finding Algorithm}
\label{sec:grpalg}

We use all groups in Tables 1, 2, and 4 of \citet{wilson16} in the lens beams to include all the observed mass.  This sample thus includes groups with as few as three member galaxies, for which group properties are difficult to accurately determine, and groups that our algorithm did not find that we added in manually with differently determined group properties.  So, the formal uncertainties on the convergence and shear (see Section \ref{sec:formalism}) consequently are large.  

To check whether this inclusiveness is biasing our results, we repeat our analysis discarding manually-added groups and others with $N_m <$ 5 (as groups with fewer members are more likely to be spurious), $\sigma_{\rm{grp}} \le$ 200 km s$^{-1}$ (such structures might be cuts through galaxy filaments or sheets misidentified as bound groups), $z_{\rm{grp}} <$ 0.1 (where only our smallest groups are sampled out to at least 1$r_{\rm{vir}}$ given our typical field sizes, resulting in possibly ill-determined group properties), or are near the field edge (as the group centroid could be biased). With this clean sample, we get qualitatively similar results.

All of our groups are at $z_{\rm{grp}} < z_{S}$, where $z_{S}$ is the redshift of the source, so any might contribute to the lensing of the source.  In most fields, our group catalog is more sensitive at redshifts below the lens redshift ($z_{L}$) than between $z_{L}$ and $z_{S}$.  \citet{mccully17} find that structures in front of the lens are more likely to significantly affect the lensing potential than those between the lens and the source.  Thus, any groups we miss at the redshifts above that of the lenses are less likely to be significant.

\subsubsection{Lens Groups}

Having characterized the groups in these 26 fields, we now see whether any of the lens galaxies are also group members, since mass at the redshift of the lens can significantly affect the lensing potential.  Our group finder does not treat lens galaxies differently, so nothing about the algorithm should be preferentially selecting structures at the lens' redshifts.  However, the fields, and thus our spectroscopic coverage, were centered on the lenses, so if there are structures they are more likely to be found than if they were at the same redshifts elsewhere in the fields.  Table \ref{table:lenssystemprops} lists the lens and lens group properties, where present, for the 26 lenses in the 26 fields in our group catalog.  As stated in \citet{wilson16}, 13 of our lenses are assigned to groups.

\begin{figure*}
\includegraphics[clip=true, width=18cm]{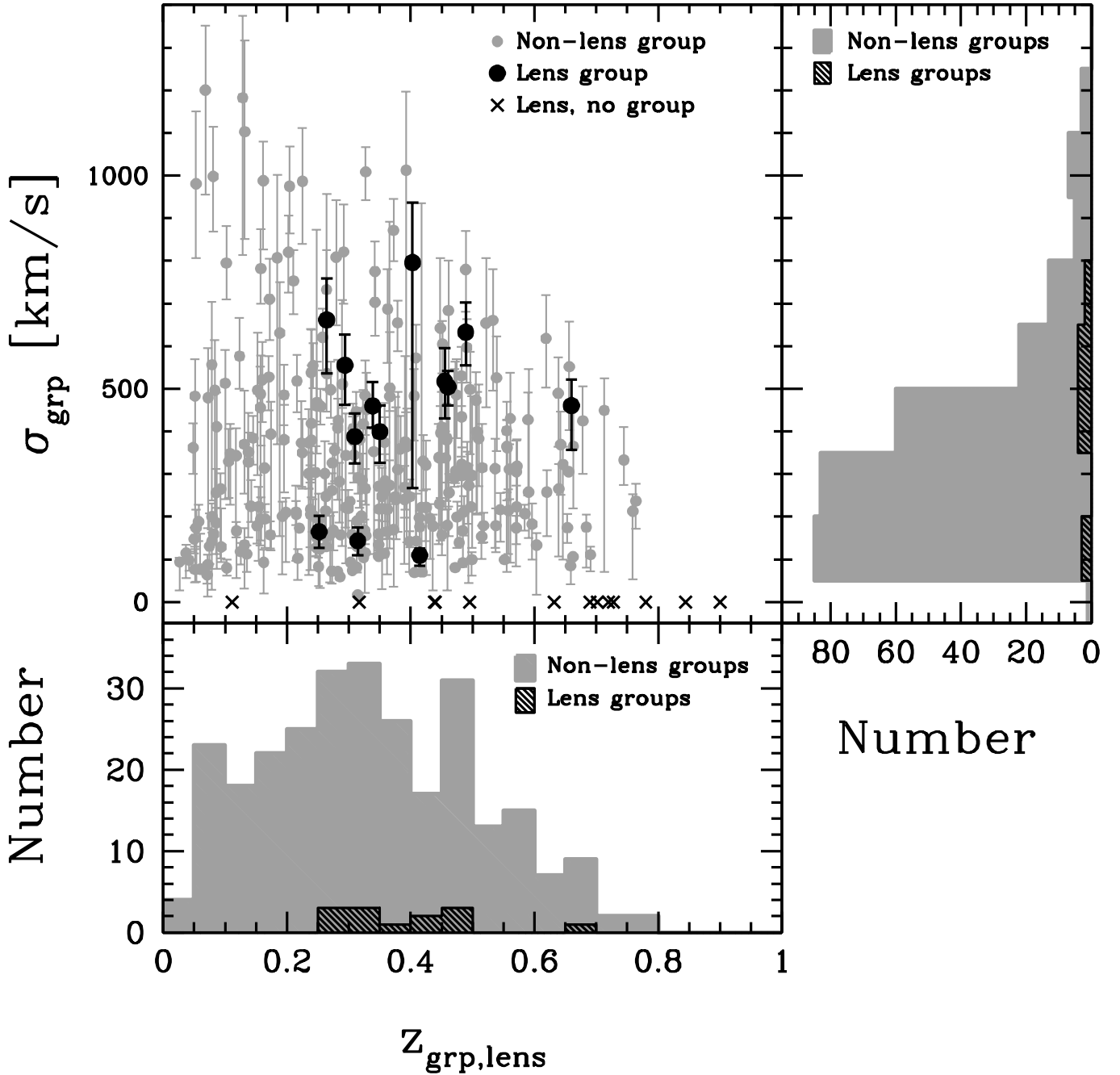}
\caption{Redshifts and velocity dispersions for our group sample (gray) with lens groups highlighted (black).  Black crosses at $\sigma_{\rm{grp}} =$ 0  mark the redshifts of lenses that are not identified as group galaxies; most are at higher redshifts where our group catalog is less sensitive.   Four lenses do not have firm spectroscopic redshifts and thus could not have been assigned group membership.  Out of 26 lenses, at least 13 are in groups.  The velocity dispersion distributions of lens and non-lens groups are statistically distinguishable using an MWW test, but this difference is likely caused by a difference in variance rather than mean.
}
\label{fig:lensingrp}
\end{figure*}

The distributions of lens group velocity dispersions and redshifts are in Figure \ref{fig:lensingrp}. The velocity dispersions range from $\sim$ 100 - 800 km s$^{-1}$ (although some have large uncertainties). Most lens groups have slightly higher velocity dispersions than the average for our overall group sample.  The effect may arise from the lower spectroscopic completeness away from the lens \citep{momcheva15}, as group velocity dispersion tends to be underestimated when its members' redshift distribution is undersampled \citep{zabludoff98}.

To ascertain if the velocity dispersions of lens groups are in fact different, we compare our lens groups only to LOS groups with projected spatial centroids within 5$^{\prime}$ of the lenses (typically the best sampled region in all our fields).  A Mann-Whitney-Wilcoxon test \citep[MWW,][]{wilcoxon45,mann47} distinguishes between the two distributions at $>$ 95\% confidence, a result likely to be dominated by the distributions' different variances.  To check for differences between the means of the two velocity dispersion distributions, we perform a ``bootstrap means comparison'' (see Appendix \ref{appendix:bootmean}), drawing from the 13 best-sampled non-lens groups and finding that their mean velocity dispersion is at least as large as that of the lens groups in 6.9\% of 1000 trials.

The falling sensitivity of our redshift catalog (Momcheva et al. 2015) results in few groups above z $\gtrsim$ 0.6.  Four of nine of the lenses with z $>$ 0.6 either have only a photometric redshift or an uncertain spectroscopic redshift, so we do not assign them group membership.  For example, \citet{auger08} identify sbs1520 as lying in a group, but we do not do so here because there are several different lens redshift estimates in the literature.

Of the other five lenses that have good spectroscopic lens redshifts, only one is classified here as a group member.  Another has a redshift nearby those of two groups, and a third lies in an apparent redshift overdensity.  MG1131 and q0158 (which is at lower redshift), have tentative group identifications based on photometry.

In summary, 12 of 17 (71\%) lenses at z $<$ 0.6 are in groups. At z $>$ 0.6, only one of five (20\%) lenses with firm spectroscopic redshifts is a group member, suggesting significant incompleteness at high $z$.  Our lenses reside in group environments at least half the time, and these groups have $\sigma_{\rm{grp}}$ comparable to others along the sightlines. We now investigate the lensing properties of these groups in more depth.

\begin{deluxetable*}{lrrllclcccccc}
\tablecaption{Lens Systems \label{table:lenssystemprops}}
\tablehead{
\colhead{Field} & \colhead{$RA_{\rm{lens}}$} & \colhead{$Dec_{\rm{lens}}$} & \colhead{$z_{L}$} & \colhead{$z_{S}$} & \colhead{Images\tablenotemark{a}} & \colhead{Size\tablenotemark{b}} & \colhead{Size type\tablenotemark{c}} & \colhead{$N_{\rm{grp}}$\tablenotemark{d}} & \colhead{$z_{\rm{grp}}$\tablenotemark{d}} & \colhead{R\tablenotemark{d,e} } & \colhead{$\sigma_{\rm{grp}}$\tablenotemark{d}} & \colhead{$r_{\rm{vir}}$\tablenotemark{d}}   \\
\colhead{} & \colhead{[deg]} & \colhead{[deg]} & \colhead{ } & \colhead{} & \colhead{} & \colhead{[$^{\prime\prime}$]} & \colhead{ } &  \colhead{ } & \colhead{ } & \colhead{[$^{\prime}$]} & \colhead{[km s$^{-1}$]}  & \colhead{[Mpc]} 
}
\startdata

    b0712 &  109.0152 &   47.1474 &  0.4060 &  1.3390 &    4 & 1.46 &  2 &        13 &  0.4030 &  1.07 &  800 $^{+   140}_{-   530}$ & 1.59 \\ 
    b1152 &  178.8264 &   19.6617 &  0.4386 &  1.0173 &    2 & 1.59 &  3 &         - &       - &     - &     -  &     - \\ 
    b1422 &  216.1587 &   22.9335 &  0.3374 &  3.6318 &   4E & 1.68 &  2 &      23 &  0.3385 &  1.34 &   460 $^{+   60}_{-   50}$ & 0.95 \\ 
    b1600 &  240.4188 &   43.2798 &  0.4140 &  1.5890 &    2 & 1.40 &  3 &      6 &  0.4146 &  0.29 &   110 $^{+   20}_{-   20}$ & 0.22 \\ 
    b2114 &  319.2116 &    2.4297 &  0.3150 &  0.5883 &  2+2 & 1.31 &  0 &        10 &  0.3143 &  0.45 &   140 $^{+   30}_{-   30}$ & 0.30 \\ 
  bri0952 &  148.7505 &   -1.5017 &  0.6320 &  4.4462 &    2 & 1.00 &  3 &         - &       - &     - &     -  &     - \\ 
  fbq0951 &  147.8441 &   26.5873 &  0.2600 &  1.2488 &    2 & 1.11 &  3 &     21 &  0.2643 &  5.81 &   660 $^{+  100}_{-  130}$ & 1.43 \\ 
    h1413 &  213.9427 &   11.4954 &  0.9\tablenotemark{f} &  2.4873 &    4 & 1.35 &  1 &         - &       - &     - &     -  &     - \\ 
   h12531 &  193.2779\tablenotemark{g} &  -29.2417\tablenotemark{g} &  0.69\tablenotemark{h}  &     -     &    4 & 1.23 &  2 &         - &       - &     - &     -  &     - \\ 
   he0435 &   69.5620 &  -12.2874 &  0.4546 &  1.6961 &    4 & 2.42 &  2 &     12 &  0.4550 &  0.61 &   520 $^{+   80}_{-   90}$ & 1.01 \\ 
   he1104 &  166.6389 &  -18.3567 &  0.7280 &  2.3207 &    2 & 3.19 &  3 &     - &       - &     - &     -  &     - \\ 
   he2149 &  328.0312 &  -27.5303 &  0.4953 &  2.0330 &    2 & 1.70 &  3 &     - &       - &     - &     -  &     - \\ 
 hst14113 &  212.8320 &   52.1916 &  0.4644 &  2.8110 &    4 & 1.80 &  2 &        55 &  0.4603 &  3.42 &   500 $^{+   40}_{-   40}$ & 0.98 \\ 
  lbq1333 &  203.8950 &    1.3015 &  0.4400 &  1.5645 &    2 & 1.63 &  3 &         - &       - &     - &     -  &     - \\ 
   mg0751 &  117.9229 &   27.2755 &  0.3502 &  3.2000 &    R & 0.70 &  0 &        26 &  0.3501 &  1.12 &   400 $^{+   60}_{-   70}$ & 0.82 \\ 
   mg1131 &  172.9853 &    4.9304 &  0.8440 &  2\tablenotemark{i} &   2R & 2.10 &  0 &         - &       - &     - &     -  &     - \\ 
   mg1549 &  237.3014 &   30.7880 &  0.1117 &  1.1700 &    R & 1.70 &  0 &         - &       - &     - &     -  &     - \\ 
   mg1654 &  253.6743 &   13.7727 &  0.2530 &  1.7400 &    R & 2.10 &  0 &         8 &  0.2520 &  1.10 &   160 $^{+   40}_{-   40}$ & 0.36 \\ 
   pg1115 &  169.5706 &    7.7663 &  0.3098 &  1.7355 &    4 & 2.32 &  2 &     13 &  0.3097 &  0.17 &   390 $^{+   50}_{-   60}$ & 0.82 \\ 
    q0047 &   12.4245 &  -27.8738 &  0.4842 &  3.5950 &  4ER & 2.70 &  0 &        20 &  0.4890 &  3.41 &   630 $^{+   70}_{-   80}$ & 1.21 \\ 
    q0158 &   29.6728 &  -43.4177 &  0.3170 &  1.2900 &    2 & 1.22 &  3 &        - &       - &     - &     -  &     - \\ 
    q1017 &  154.3497 &  -20.7829 &  0.78\tablenotemark{h} &  2.5450 &    2 & 0.85 &  3 &         - &       - &     - &     -  &     - \\ 
    q1355 &  208.9308 &  -22.9564 &  0.7020 &  1.37\tablenotemark{g} &    2 & 1.23 &  3 &         - &       - &     - &     -  &     - \\ 
  rxj1131 &  172.9646 &  -12.5329 &  0.2950 &  0.6580 &    4 & 3.80 &  2 &     38 &  0.2938 &  5.23 &   550 $^{+   70}_{-   90}$ & 1.18 \\ 
  sbs1520 &  230.4368 &   52.9135 &  0.72,\tablenotemark{j}0.761\tablenotemark{k} &  1.86\tablenotemark{g} &    2 & 1.59 &  3 &     - &       - &     - &     -  &     - \\ 
  wfi2033 &  308.4247 &  -47.3952 &  0.6610 &  1.6604 &    4 & 2.33 &  2 &        14 &  0.6598 &  1.94 &   460 $^{+   60}_{-  100}$ & 0.80 \\ 

\enddata
 
\begin{minipage}{12cm}
\tablenotetext{a}{From CASTLES (https://www.cfa.harvard.edu/castles/): Number = Number of images, R = Einstein ring, and E = Extended.}
\tablenotetext{b}{Estimates of twice the average Einstein radius from CASTLES.}
\tablenotetext{c}{From CASTLES: 0 = Literature estimate, 1 = Maximum image pair separation, 2 = Twice average distance from lens center to images,  3 = Twice an SIS model's critical radius plus external shear.}
\tablenotetext{d}{Lens group properties from \citet{wilson16}.}
\tablenotetext{e}{Projected angular distance between the lens and the group centroid.}
\tablenotetext{f}{Photometric redshift from \citet{kneib98photz}.}
\tablenotetext{g}{From CASTLES.}
\tablenotetext{h}{Photometric redshift from CASTLES.}
\tablenotetext{i}{Photometric redshift from \citet{kochanek00}.}
\tablenotetext{j}{Spectroscopic redshift from \citet{chavushyan97}.}
\tablenotetext{k}{Spectroscopic redshift from \citet{auger08}.}

\end{minipage}

\end{deluxetable*}

\section{Lensing Formalism}
\label{sec:formalism}

First, we summarize how convergence and shear enter into the calculation of $H_0$ from time-delay measurements \citep[e.g. Suyu et al. 2013; for a review of time delay cosmography, see][]{treu16}.

The extra time it takes light from the source to reach the observer compared to if there was no lensing is
\begin{equation}
\label{equn:t}
t(\pmb{\theta},\pmb{\beta}) = \frac{D_{\Delta t}}{c} \left[\frac{\left(\pmb{\theta} - \pmb{\beta}\right)^2}{2} - \psi (\pmb{\theta})\right],
\end{equation}
where image angular position $\pmb{\theta} = (\theta_1, \theta_2)$, source position $\pmb{\beta} = (\beta_1, \beta_2)$, $c$ is the speed of light, and $\psi(\pmb{\theta})$ is the lens potential.  $D_{\Delta t}$ is the time-delay distance,
\begin{equation}
D_{\Delta t} \equiv (1+z_L)\frac{D_L D_S}{D_{LS}},
\end{equation}
where $z_L$ is the lens redshift and $D_L$, $D_S$, and $D_{LS}$ are angular diameter distances between, respectively, the observer and the lens, the observer and the source, and between the source and the lens.  The time-delay distance is inversely proportional to $H_0$, as well as weakly dependent on other cosmological parameters.

Because of the mass sheet degeneracy, lensing data alone cannot constrain external convergence.  Therefore it is common to fit models that omit convergence, which yield an estimate for the time delay distance of $D^{\rm{model}}_{\Delta t}$.  When external convergence is applied, the new estimate for $D_{\Delta t}$ is

\begin{equation}
D_{\Delta t} = \frac{D^{\rm{model}}_{\Delta t}}{1 - \kappa}.
\end{equation}

Convergences thus enter the calculation as 1 - $\kappa$ from the mass sheet degeneracy, and shear appears inside the model-dependent factor.

We calculate the effective convergence ($\kappa_{\rm eff}$) and shear ($\gamma_{\rm eff}$) for each group, as well as total convergence ($\kappa_{\rm{tot}}$) and shear ($\gamma_{\rm{tot}}$) for each field, using the following method.

Here we focus on determining the importance of \textit{groups} to the lensing potential.  We do not model the lens galaxy.  We also do not consider any galaxies that have not been assigned to groups, leaving that to more detailed future modeling \cite[e.g.,][]{wong11,mccully17}.  

We assume that each group's mass is described by a group halo and that there is no appreciable mass bound to the individual group galaxies (the group halo limit).  Using preliminary redshift catalogs for eight of these systems, \citet{momcheva06} find that assigning the mass to the group halo or to individual member galaxy halos does not affect which groups are significant to the lensing potential.

Using weak lensing measurements of galaxy groups with masses $\sim 10^{13} - 10^{14.5} M_{\odot}$, \citet{viola15} find that the density profiles agree with that expected for Navarro, Frenk, and White (NFW) profiles \citep{navarro96}. So, we model the groups in each lens field as NFW halos, estimating the scale radius $r_s$ and central density $\rho_s$ of a halo from its redshift and velocity dispersion using the results of simulations by \citet{zhao09}.  We assume the groups generally are virialized, as many have centrally concentrated early type populations.  For any non-virialized systems, the masses derived from the velocity dispersions here could be overestimated by up to $\sim 2\times$, if the measured dispersion is closer to the infall velocity.  This systematic would bias the lensing contribution upward.

We then calculate the convergence ($\kappa$) and shear ($\gamma$) for each halo using the truncated NFW profile formalism given in \citet{baltz09} to calculate the projected surface mass density ($\Sigma(x)$) and the mean projected surface density ($\bar\Sigma$).  For these, we use $x = r/r_s$, where $r$ is the distance between the group center and the lens LOS at the redshift of the group, and $\tau = \frac{r_t}{r_s}$, where $r_t$ is the truncation radius.  We choose $r_{\rm{200m}}$ as our truncation radius  (see Appendix \ref{appendix:truncationchoice} for a discussion).  We calculate 
\begin{equation}
\kappa_s = \frac{r_s\rho_s}{\Sigma_{\rm{crit}}},
\end{equation}
and
\begin{equation}
\Sigma_{\rm{crit}} = \frac{c^2}{4\pi{G}} \frac{D_S}{D_PD_{PS}},
\end{equation}
where $D_{S}$, $D_P$, and $D_{PS}$ are the angular diameter distances between the observer and the source, the observer and the perturbing group, and the perturbing group and the source, respectively.  Next, we calculate the convergence
\begin{equation}
\kappa = \frac{\Sigma}{\Sigma_{\rm{crit}}}
\end{equation}
and the shear
\begin{equation}
\gamma = \frac{\bar\Sigma-\Sigma}{\Sigma_{\rm{crit}}}
\end{equation}
in the perturber's plane.

Now, as in \citet{momcheva06}, we define
\begin{equation}
\beta = \frac{D(z_1,z_2) D(0,z_S)}{D(0,z_2) D(z_1,z_S)},
\end{equation}
where
\begin{equation}
z_1 = \text{min}(z_L, z_{P}),
\end{equation}
and
\begin{equation}
z_2 = \text{max}(z_L,z_{P}).
\end{equation}

When the convergence and shear is due to a perturber, the expressions for its effective convergence and shear more generally are 
\begin{equation}
\kappa_{\rm eff} = \frac{ \left(1-\beta\right)\left[\kappa - \beta \left(\kappa^2-\gamma^2\right)\right] }{  \left(1-\beta\kappa \right)^2-\left(\beta\gamma \right)^2  }
\end{equation}
and
\begin{equation}
\gamma_{\rm eff} = \frac{ \left(1-\beta \right)\gamma }{ \left(1-\beta\kappa \right)^2-\left(\beta\gamma \right)^2 },
\end{equation}
as given in \citet{keeton03}.  In this analysis, all groups are treated the same, regardless of whether the lens galaxy is a member.  For the groups that do have lens galaxies as members, $\beta$ is small, approaching zero for a lens that is near the velocity centroid of its group, so $\kappa_{\rm eff}$ and $\gamma_{\rm eff}$ will approach $\kappa$ and $\gamma$.

We estimate the total effective convergence due to all the groups in the field for each lens LOS using
\begin{equation}
\kappa_{\rm{tot}} = \sum_{i} \kappa_{\rm eff,i}
\end{equation}
from \citet{momcheva06}.  We calculate the effective convergence due to each group halo, including the lens group halo, and sum them in each field to estimate the importance of LOS structures and lens environment for the lensing potential.  

Since the shears add as tensors, we measure $\theta_{\gamma}$, the position angle measured north through east between the group projected spatial centroid and the lens.  This definition is consistent with the direction in which the lensing critical curve is stretched but is orthogonal to the direction in which an image would be stretched.

We then calculate the total shear components for each field as
\begin{equation}
\gamma_{c,tot} = \sum_{i} \gamma_{\rm eff} \cos 2\theta_{\gamma,i},
\end{equation}
\begin{equation}
\gamma_{s,tot} = \sum_{i} \gamma_{\rm eff} \sin 2\theta_{\gamma,i},
\end{equation}
and
\begin{equation}
\gamma_{\rm{tot}} = \sqrt{\gamma_{c,\rm{tot}}^2 + \gamma_{s,\rm{tot}}^2}.%,
\end{equation}
The total position angle of the shear is
\begin{equation}
\theta_{\gamma,\rm{tot}} = \frac{1}{2}\arctan{\left(\frac{\gamma_{s,\rm{tot}}}{\gamma_{c,\rm{tot}}}\right)}.
\end{equation}

These calculations for the total values assume that there are no lensing interactions between the halos which arise from cross terms in the multi-plane lens equation, which is a simplification \cite[see][]{mccully14}.

We may underestimate the velocity dispersions of groups at $z \gtrsim$ 0.6, because a lower fraction of true members may have been observed due to our lower sensitivity.  This underestimate would result in a underestimation of their effect on the lensing.  Thus, the true properties would result in stronger effects, making our results conservative.

Plots showing the distribution of the angular and redshift separation of groups from the lens color coded by their $\kappa_{\rm eff}$ are in Figure \ref{fig:loslensingeffectskappa}.  Sky plots of the fields with groups marked and their $\gamma_{\rm eff}$'s indicated are in Figure \ref{fig:gammavecfield}.  Groups contribute to the total convergence in several ways.  Some fields have one dominant group, some have multiple important ones, and one has several unimportant groups that add up to a significant contribution.  In the cases where there is one dominant group, this group is not always the lens group.  As shears can add or subtract depending on relative position angles, fields with multiple important groups can have larger or smaller $\gamma_{\rm{tot}}$.

\begin{figure*}
\includegraphics[clip=true, width=18cm]{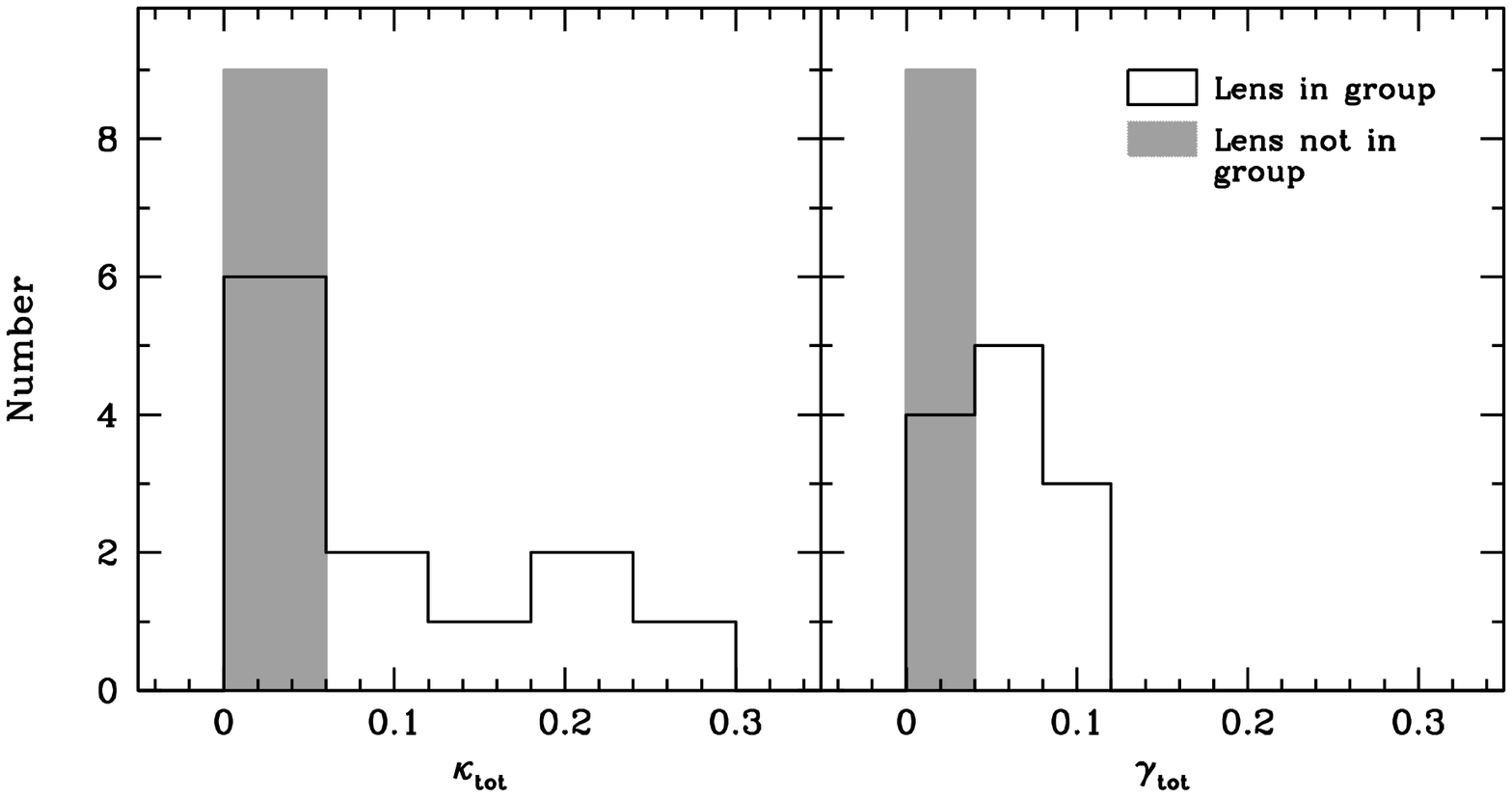}
\caption{Distribution of $\kappa_{\rm{tot}}$ (\textit{left}) and  $\gamma_{\rm{tot}}$ (\textit{right}) for fields with lenses in groups (white histograms) and those with lenses that have not been identified as group members (filled gray) for the 21 fields with firm spectroscopic lens redshifts and without a supergroup.  These total values include the lens group contributions.  Bin widths are different for $\kappa_{\rm{tot}}$ and $\gamma_{\rm{tot}}$ because of their different average uncertainties.  Fields with lens groups have a tail extending to large $\kappa_{tot}$ that is not present in fields without lens groups.
}
\label{fig:totvalslensingrpvsnot}
\end{figure*}

Figure \ref{fig:totvalslensingrpvsnot} shows the histograms of $\kappa_{tot}$ and $\gamma_{\rm{tot}}$ for fields with and without lens groups.  A Kolmogorov-Smirnov (K-S) test comparing the unbinned distributions is significant at 95\% confidence, suggesting that there is a tail to higher $\kappa_{tot}$ and $\gamma_{\rm{tot}}$ in fields where the lens lies in a group.  Furthermore, our bootstrap means comparison (Appendix \ref{appendix:bootmean}) indicates a $<$ 0.1\% chance that the mean $\kappa_{\rm tot}$ and $\gamma_{\rm tot}$ for fields with lenses in groups are as small as the values for fields with lenses not in groups.  Table \ref{table:summary} summarizes the lensing parameters for each field, and Table \ref{table:totlensmax} lists the individual group properties and lensing parameters in each field.

All statistical tests are performed on unbinned data; we bin only to visually present the data.

\begin{figure*}
\includegraphics[clip=true, width=18cm]{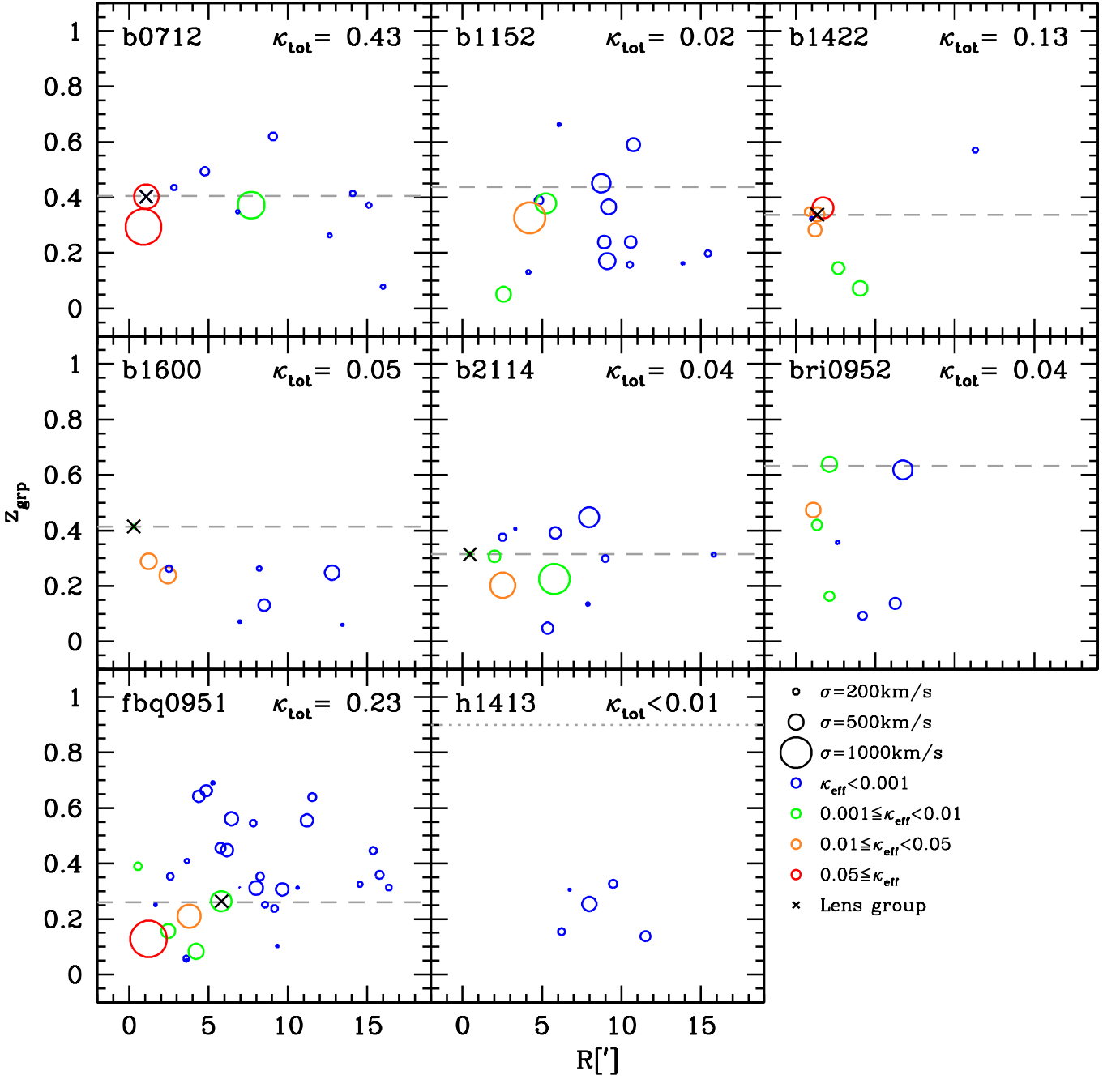}
\caption{Redshift, angular separation from the lens, and convergence values of groups (open circles) in our 25 fields with lens and source redshift measurements.  Circle sizes are scaled by group velocity dispersion.  Points are color coded by $\kappa_{\rm eff}$.  Groups that include the lens galaxies are marked additionally with black crosses.  The supergroup in field b0712 is included.  Gray dashed lines mark spectroscopic lens redshifts, and gray dotted lines mark photometric redshifts.  The lens of sbs1520 has two possible values, so we use $z_{L}$ = 0.72 in the calculations but mark where the other value would be with a second dashed gray line.  Groups behind the lens are above the gray lines, and those in the foreground are below.  Red or orange circles indicate significant groups, usually because they are near the lens in redshift and/or on the sky (small R) and/or are massive (large circles).  Groups marked in blue or green are not important, usually because they are far from the lens and/or are low mass.  There is a diversity in how groups contribute to their fields' lensing potentials.  Some have one clearly dominant group, some have multiple important groups, and one has many individually insignificant groups that add up to an overall significant $\kappa_{\rm{tot}}$. }
\label{fig:loslensingeffectskappa}
\end{figure*}
\begin{figure*}
\ContinuedFloat
\includegraphics[clip=true, width=18cm]{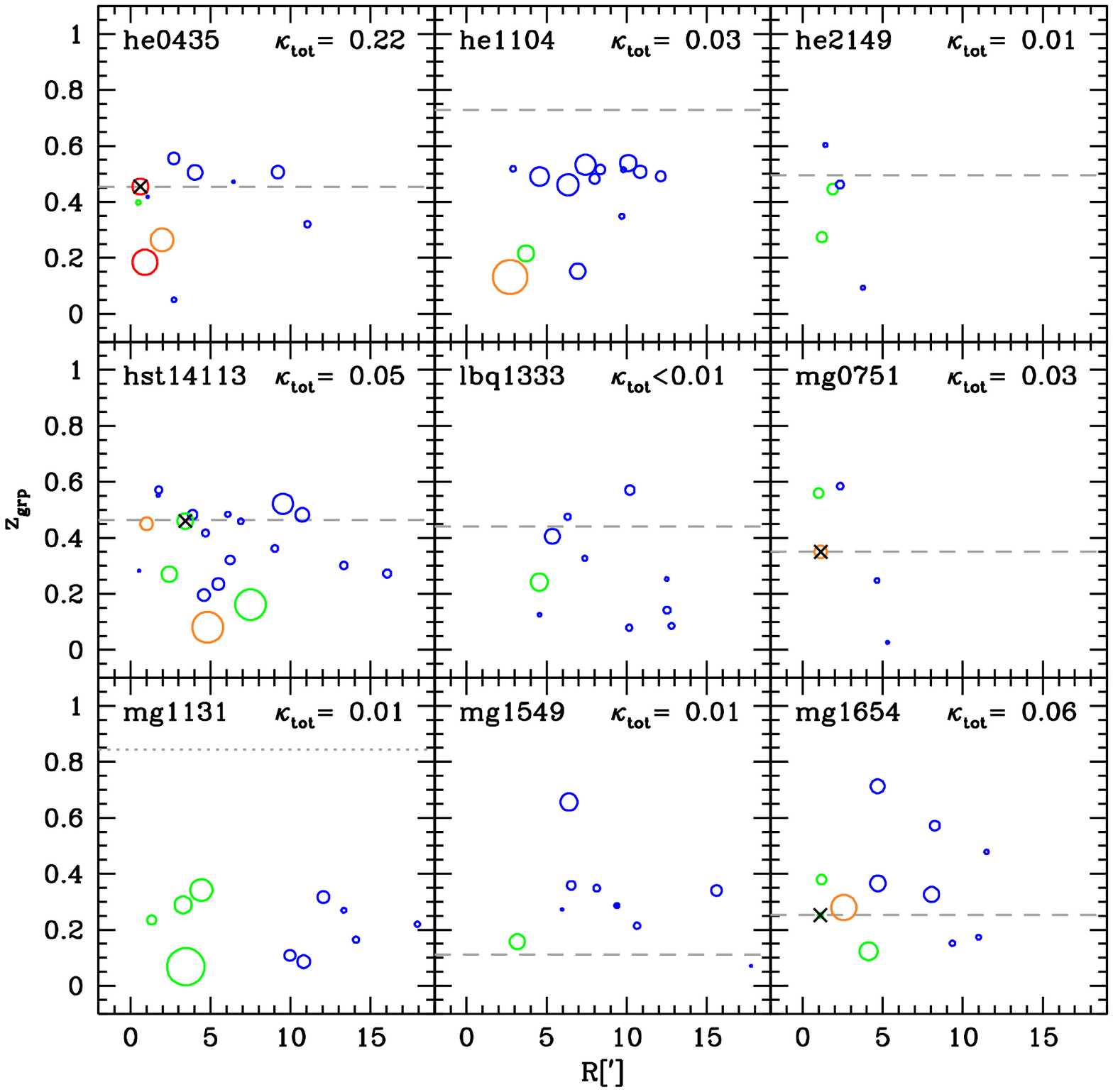}
\caption{Continued.}
\end{figure*}
\begin{figure*}
\ContinuedFloat
\includegraphics[clip=true, width=18cm]{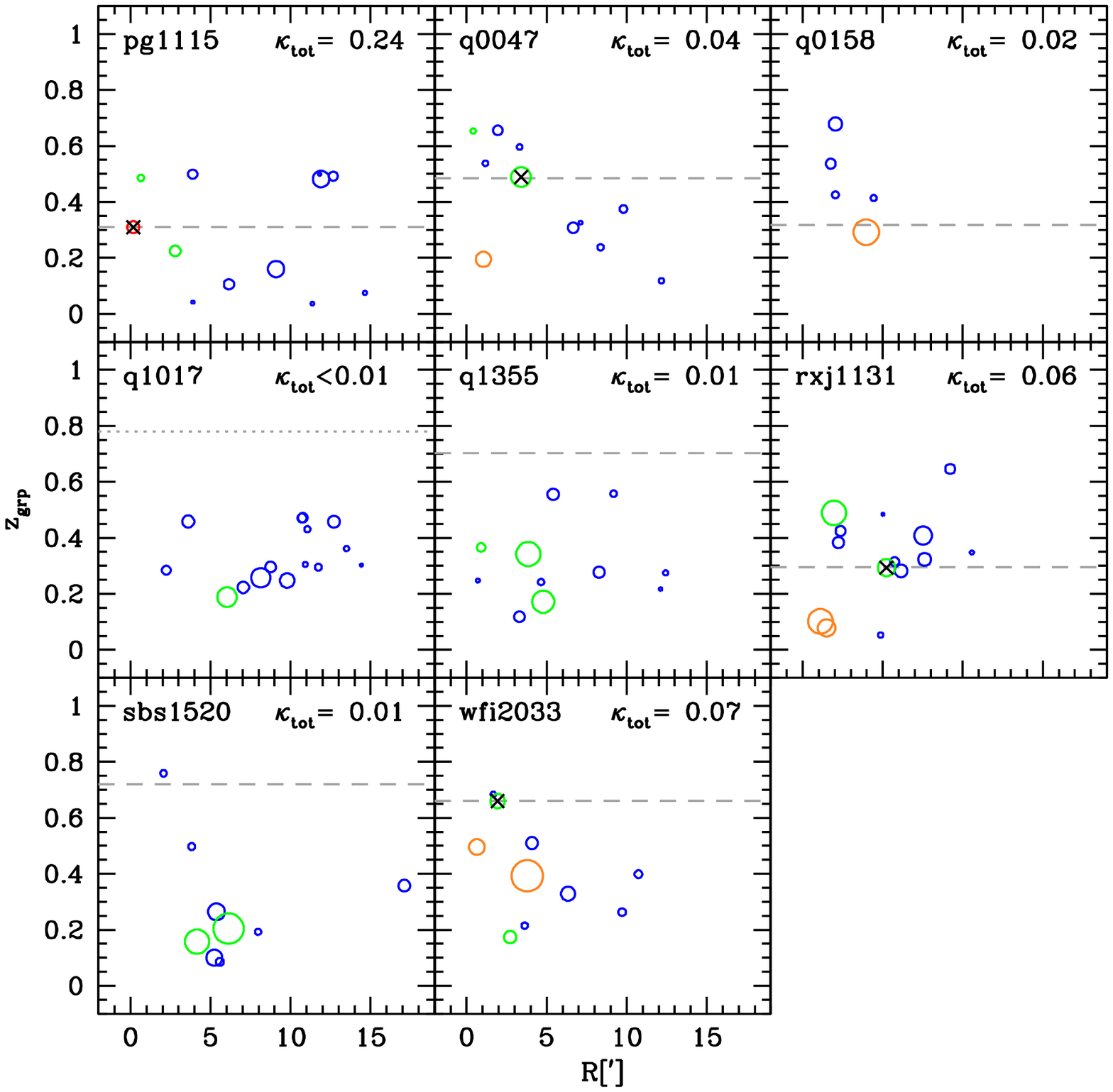}
\caption{Continued.}
\end{figure*}

\begin{figure*}
\includegraphics[clip=true, width=18cm]{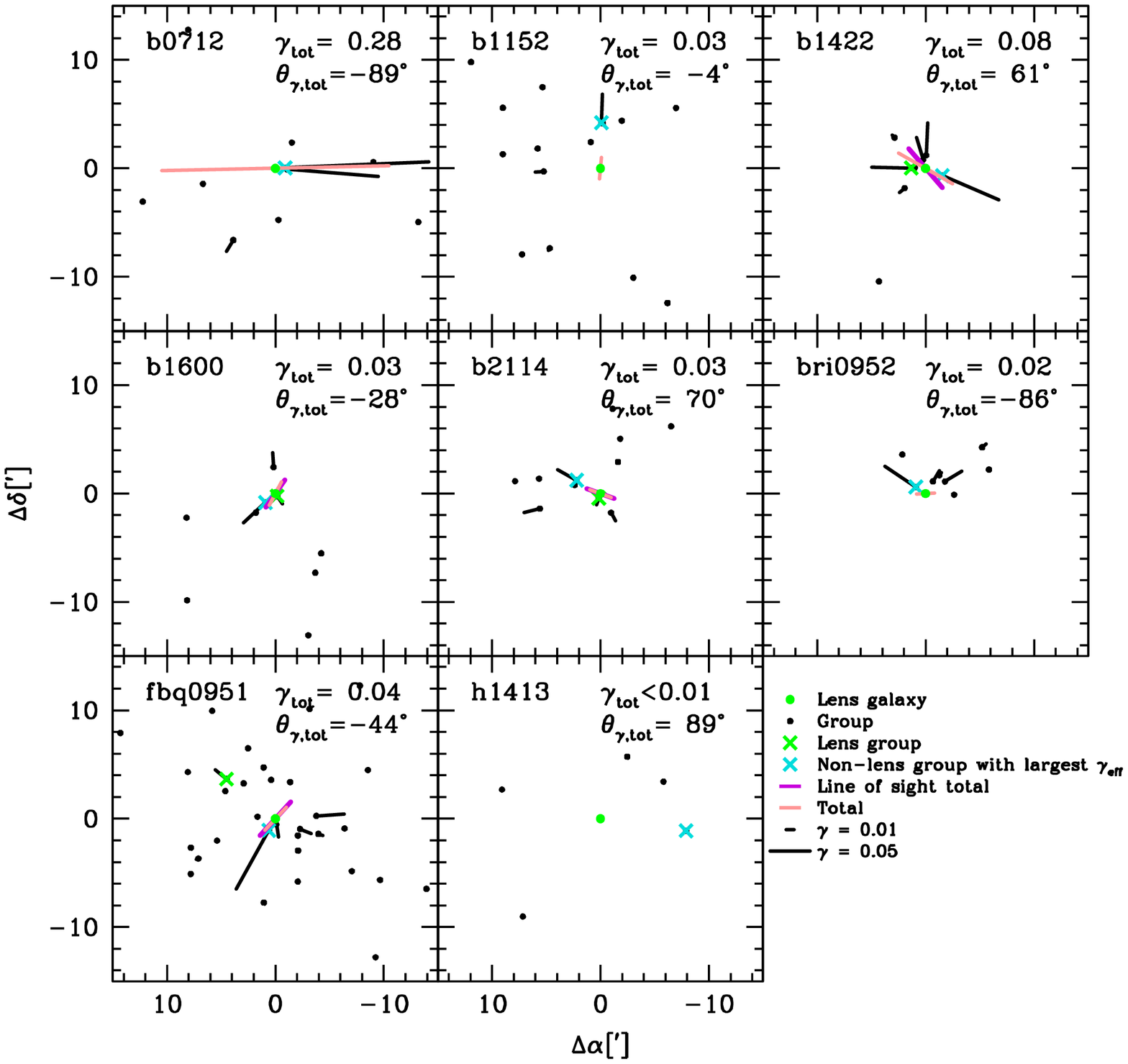}
\caption{The distribution of groups on the sky relative to the lens for our 25 fields with lens and source redshifts.  Black points mark groups, green crosses mark lens groups, cyan crosses mark the LOS groups with the largest $\gamma_{\rm eff}$, and the green filled circles mark lens galaxies.  The supergroup in field b0712 is included.  Black lines are scaled by the group's $\gamma_{\rm eff}$ and point away from the lens to depict the group's position angle, although the shear is invariant under a 180$^{\circ}$ rotation.  Groups without lines have shears too small for the line to be visible ($\gamma_{\rm eff} \lesssim$ 0.003).  The field's $\gamma_{\rm{tot}}$ is represented by the pink line, and the field's $\gamma_{\rm{los}}$, the total shear without the lens galaxy, by the purple line for fields with lens groups.  These values also are invariant under a 180$^{\circ}$ rotation and are scaled so the full length is equal to the line length of a group with equal $\gamma_{\rm eff}$.  Since we define the position angle to point in the direction of the perturber, the critical curves would be elongated in the directions represented by these lines, and images would be stretched orthogonally. The typical uncertainties in $\gamma_{\rm eff}$ and $\gamma_{\rm{tot}}$ are, respectively, 0.003 and 0.02.  Although we represent the group shears as a vector field they add as tensors.  For some fields, the direction and magnitude of $\gamma_{\rm{tot}}$ is driven by one group. This group is not always the lens group.  Several fields have several significant groups; their shears can add or subtract depending on their relative position angles.}
\label{fig:gammavecfield}
\end{figure*}
\begin{figure*}
\ContinuedFloat
\includegraphics[clip=true, width=18cm]{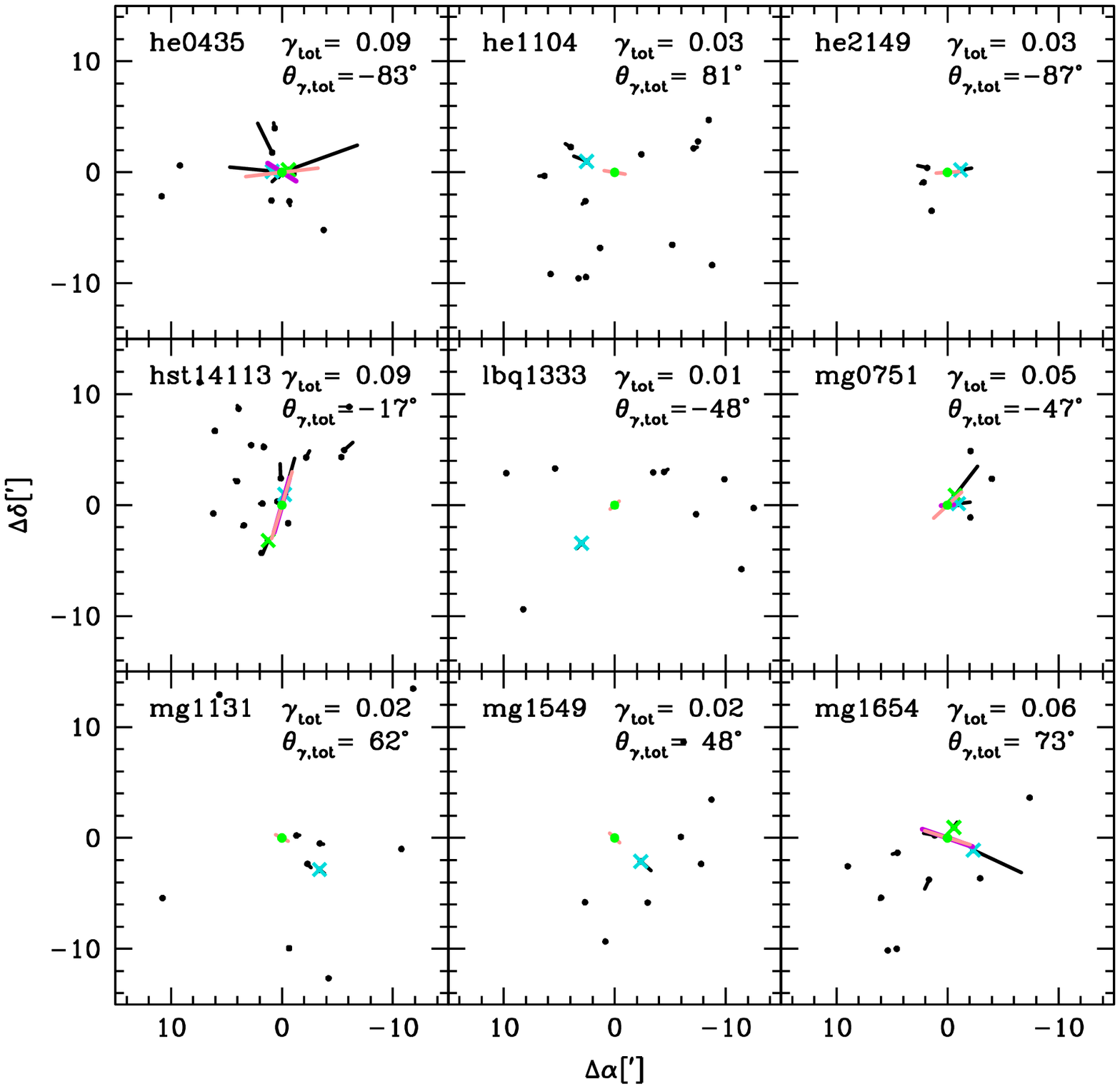}
\caption{Continued.}
\end{figure*}
\begin{figure*}
\ContinuedFloat
\includegraphics[clip=true, width=18cm]{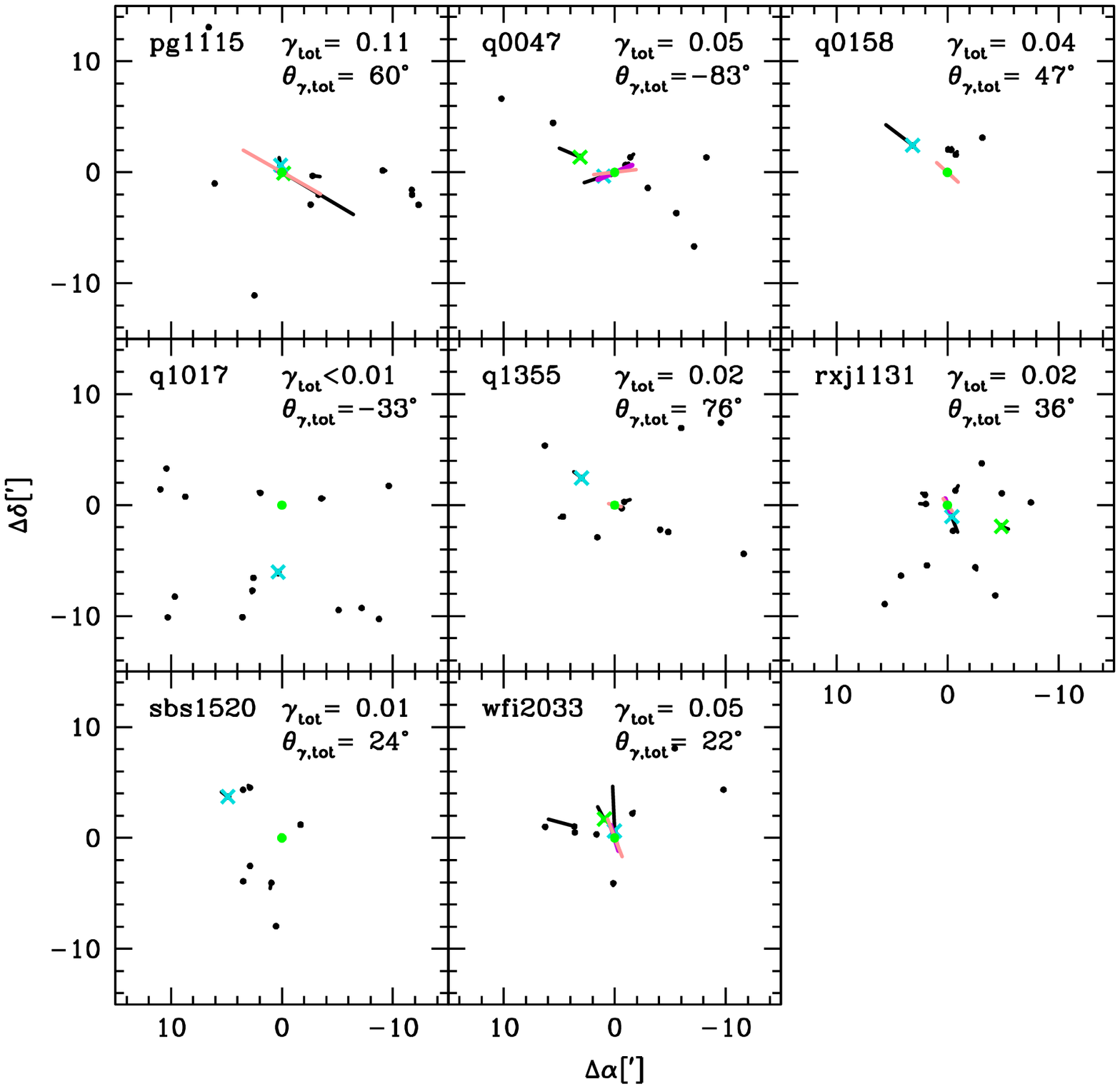}
\caption{Continued.}
\end{figure*}

\begin{deluxetable*}{lrllllll}
\tablecaption{Summary of Lensing Parameters for Each Field \label{table:summary}}
\tablehead{
\colhead{Field} & \colhead{$N_{\rm{LOS grps}}$} & \colhead{$\kappa_{\rm{lens}}$} & \colhead{$\gamma_{\rm{lens}}$} & \colhead{$\kappa_{\rm{los}}$} & \colhead{$\gamma_{\rm{los}}$} & \colhead{$\kappa_{\rm{tot}}$} & \colhead{$\gamma_{\rm{tot}}$} 
}
\startdata

    b0712\tablenotemark{a} & 11 & 0.118 $^{+  0.050}_{-  0.014}$ &   0.115 $^{+  0.047}_{-  0.038}$ &   0.311 $^{+  0.229}_{ -0.116}$ &   0.173 $^{+  0.076}_{-  0.048}$ &   0.429 $^{+  0.225}_{-  0.126}$ &   0.284 $^{+  0.087}_{-  0.058}$ \\ 
    b1152 & 15 & - & - &   0.021 $^{+  0.010}_{-  0.006}$ &   0.027 $^{+  0.010}_{-  0.007}$ &   0.021 $^{+  0.010}_{-  0.006}$ &   0.027 $^{+  0.010}_{-  0.007}$ \\ 
    b1422 & 8  &   0.029 $^{+  0.022}_{-  0.012}$ &   0.050 $^{+  0.019}_{-  0.014}$ &   0.102 $^{+  0.088}_{ -0.037}$ &   0.065 $^{+  0.051}_{ -0.034}$ &   0.130 $^{+  0.084}_{-  0.044}$ &   0.078 $^{+  0.050}_{-  0.043}$ \\ 
    b1600 & 9  &   0.005 $^{+  0.015}_{-  0.003}$ &   0.012 $^{+  0.014}_{-  0.006}$ &   0.041 $^{+  0.025}_{ -0.012}$ &   0.041 $^{+  0.017}_{ -0.014}$ &   0.046 $^{+  0.042}_{-  0.016}$ &   0.033 $^{+  0.022}_{-  0.015}$ \\ 
    b2114 & 12 &   0.004 $^{+  0.005}_{-  0.001}$ &   0.009 $^{+  0.006}_{-  0.003}$ &   0.038 $^{+  0.025}_{ -0.011}$ &   0.037 $^{+  0.016}_{ -0.012}$ &   0.042 $^{+  0.027}_{-  0.012}$ &   0.028 $^{+  0.016}_{-  0.014}$ \\ 
  bri0952 & 8  & - & - &   0.039 $^{+  0.044}_{-  0.018}$ &   0.023 $^{+  0.035}_{-  0.023}$ &   0.039 $^{+  0.044}_{-  0.018}$ &   0.023 $^{+  0.035}_{-  0.023}$ \\ 
  fbq0951 & 32 &   0.004 $^{+  0.003}_{-  0.002}$ &   0.018 $^{+  0.007}_{-  0.007}$ &   0.226 $^{+  0.103}_{ -0.060}$ &   0.057 $^{+  0.042}_{ -0.028}$ &   0.230 $^{+  0.106}_{-  0.060}$ &   0.039 $^{+  0.045}_{-  0.028}$ \\  
    h1413 & 5  & - & - &   -                          &   0.001 $^{+  0.001}_{-  0.0004}$ &   -                          &   0.001 $^{+  0.001}_{-  0.0004}$ \\ 
   he0435 & 11 &   0.091 $^{+  0.110}_{-  0.031}$ &   0.090 $^{+  0.049}_{-  0.030}$ &   0.128 $^{+  0.063}_{ -0.029}$ &   0.041 $^{+  0.031}_{ -0.020}$ &   0.219 $^{+  0.138}_{-  0.062}$ &   0.089 $^{+  0.052}_{-  0.040}$ \\ 
   he1104 & 14 & - & - &   0.027 $^{+  0.029}_{-  0.010}$ &   0.026 $^{+  0.015}_{-  0.009}$ &   0.027 $^{+  0.029}_{-  0.010}$ &   0.026 $^{+  0.015}_{-  0.009}$ \\ 
   he2149 & 5  & - & - &   0.010 $^{+  0.015}_{-  0.006}$ &   0.028 $^{+  0.013}_{-  0.008}$ &   0.010 $^{+  0.015}_{-  0.006}$ &   0.028 $^{+  0.013}_{-  0.008}$ \\ 
 hst14113 & 20 &   0.002 $^{+  0.002}_{-  0.001}$ &   0.016 $^{+  0.007}_{-  0.005}$ &   0.052 $^{+  0.030}_{ -0.020}$ &   0.070 $^{+  0.032}_{ -0.030}$ &   0.054 $^{+  0.030}_{-  0.019}$ &   0.085 $^{+  0.035}_{-  0.030}$ \\ 
  lbq1333 & 10 & - & - &   0.003 $^{+  0.002}_{-  0.001}$ &   0.014 $^{+  0.004}_{-  0.004}$ &   0.003 $^{+  0.002}_{-  0.001}$ &   0.014 $^{+  0.004}_{-  0.004}$ \\ 
   mg0751 & 5  &   0.025 $^{+  0.022}_{-  0.011}$ &   0.044 $^{+  0.019}_{-  0.015}$ &   0.005 $^{+  0.010}_{ -0.002}$ &   0.015 $^{+  0.013}_{ -0.007}$ &   0.030 $^{+  0.027}_{-  0.014}$ &   0.046 $^{+  0.022}_{-  0.018}$ \\ 
   mg1131 & 10 & - & - &   0.014 $^{+  0.008}_{-  0.005}$ &   0.017 $^{+  0.005}_{-  0.005}$ &   0.014 $^{+  0.008}_{-  0.005}$ &   0.017 $^{+  0.005}_{-  0.005}$ \\ 
   mg1549 & 10 & - & - &   0.009 $^{+  0.016}_{-  0.004}$ &   0.017 $^{+  0.014}_{-  0.008}$ &   0.009 $^{+  0.016}_{-  0.004}$ &   0.017 $^{+  0.014}_{-  0.008}$ \\ 
   mg1654 & 11 &   0.002 $^{+  0.005}_{-  0.001}$ &   0.008 $^{+  0.009}_{-  0.004}$ &   0.058 $^{+  0.065}_{ -0.029}$ &   0.065 $^{+  0.044}_{ -0.030}$ &   0.060 $^{+  0.065}_{-  0.030}$ &   0.058 $^{+  0.046}_{-  0.029}$ \\ 
   pg1115 & 12 &   0.234 $^{+  0.102}_{-  0.059}$ &   0.099 $^{+  0.021}_{-  0.025}$ &   0.007 $^{+  0.005}_{ -0.003}$ &   0.009 $^{+  0.007}_{ -0.006}$ &   0.241 $^{+  0.101}_{-  0.059}$ &   0.108 $^{+  0.022}_{-  0.027}$ \\ 
    q0047 & 11 &   0.005 $^{+  0.007}_{-  0.003}$ &   0.027 $^{+  0.017}_{-  0.011}$ &   0.033 $^{+  0.028}_{ -0.011}$ &   0.046 $^{+  0.015}_{ -0.010}$ &   0.039 $^{+  0.031}_{-  0.014}$ &   0.052 $^{+  0.020}_{-  0.016}$ \\ 
    q0158 & 5  & - & - &   0.018 $^{+  0.026}_{-  0.010}$ &   0.035 $^{+  0.028}_{-  0.017}$ &   0.018 $^{+  0.026}_{-  0.010}$ &   0.035 $^{+  0.028}_{-  0.017}$ \\ 
    q1017 & 15 & - & - &   0.003 $^{+  0.003}_{-  0.001}$ &   0.002 $^{+  0.003}_{-  0.002}$ &   0.003 $^{+  0.003}_{-  0.001}$ &   0.002 $^{+  0.003}_{-  0.002}$ \\ 
    q1355 & 11 & - & - &   0.011 $^{+  0.011}_{-  0.004}$ &   0.016 $^{+  0.010}_{-  0.006}$ &   0.011 $^{+  0.011}_{-  0.004}$ &   0.016 $^{+  0.010}_{-  0.006}$ \\ 
  rxj1131 & 14 &   0.002 $^{+  0.002}_{-  0.001}$ &   0.009 $^{+  0.005}_{-  0.004}$ &   0.061 $^{+  0.024}_{ -0.016}$ &   0.017 $^{+  0.009}_{ -0.008}$ &   0.063 $^{+  0.024}_{-  0.015}$ &   0.019 $^{+  0.012}_{-  0.010}$ \\ 
  sbs1520 & 9  & - & - &   0.012 $^{+  0.006}_{-  0.004}$ &   0.008 $^{+  0.006}_{-  0.004}$ &   0.012 $^{+  0.006}_{-  0.004}$ &   0.008 $^{+  0.006}_{-  0.004}$ \\ 
  wfi2033 & 10 &   0.003 $^{+  0.002}_{-  0.002}$ &   0.017 $^{+  0.007}_{-  0.008}$ &   0.065 $^{+  0.053}_{ -0.021}$ &   0.033 $^{+  0.034}_{ -0.022}$ &   0.068 $^{+  0.053}_{-  0.022}$ &   0.049 $^{+  0.030}_{-  0.024}$ \\ 

\enddata
 
\begin{minipage}{12cm}
\tablenotetext{a}{Values presented here include the supergroup.  See Section \ref{sec:formalism} and Table \ref{table:totlensmax} for details.}

\end{minipage}

\end{deluxetable*}

In these plots and tables, we include all 25 fields with spectroscopic or photometric lens and source redshift estimates.  Field h12531 is discarded because it lacks a source redshift estimate.  The fields surrounding four lenses without firm spectroscopic redshifts (h1413, h12531, q1017, and sbs1520) could not have lens groups identified, so for the rest of the analysis we discard them.

In field b0712, our group finding algorithm identifies a supergroup \citep[a structure with multiple well-populated clumps in both velocity and projected on the sky;][]{wilson16}.  We separately calculate the convergence and shear of the supergroup as a whole and for its three main substructures, as listed in Table 3 of \citet{wilson16}.  If the supergroup is treated as one monolithic halo, it is very significant ($\kappa_{\rm eff} =$ 0.308$^{+0.229}_{-0.115}$, $\gamma_{\rm eff} =$ 0.180$^{+0.072}_{-0.042}$).  If the mass distribution is better described by treating the substructures separately, none have $\kappa_{\rm eff}$ or $\gamma_{\rm eff} \ge$ 0.01.  Accurately determining how big an effect this supergroup has on the lensing potential requires further observations and analysis.  We thus include this field in our field overview plots (Figures \ref{fig:loslensingeffectskappa} and \ref{fig:gammavecfield}) and in our discussion of lens group environments (Section \ref{sec:lensenvirons}), but we discard it elsewhere because of these uncertainties.

We calculate uncertainties in our effective convergences and shears due to the uncertainty in the measured group properties using the bootstrap method.  We re-sample with replacement the identified group members and recalculate $z_{\rm{grp}}$, $\sigma_{\rm{grp}}$, and the group projected spatial centroid for 1000 iterations.  Using those recalculated group properties, we then calculate the resulting $\kappa_{\rm eff}$ and $\gamma_{\rm eff}$'s.  We then calculate and subtract the median $\kappa_{\rm eff}$ and $\gamma_{\rm eff}$ values over all 1000 iterations and determine the 16th and 84th percentiles for individual groups as well as for $\kappa_{\rm{tot}}$ and $\gamma_{\rm{tot}}$.  

The median values can differ from the measured values.  Since we resample the observed group members with replacement, a bootstrap realization's velocity dispersion is more likely to be smaller than measured, leading to smaller convergences and shears.  The fractional differences in the measured and median values (i.e., $(\kappa_{\rm eff,obs}-\kappa_{\rm eff,med})/\kappa_{\rm eff,obs}$) show no strong trend with increasing lensing parameter value.  Also, the mean fractional errors in the lensing parameters are almost twice as large as the mean fractional differences due to the bootstrap realizations (a fractional difference of the median from the observed convergence of 0.37 versus a fractional differences of the error in convergence from the observed convergence of 0.60, and 0.28 versus 0.50 for corresponding quantities for shear).

We do not incorporate uncertainty in the NFW concentration parameter associated with scatter in the mass/concentration relation.  \citet{wong11} find that such uncertainties affect shears on the order of $\lesssim$ 0.01, generally when the group centroid is very close to the lens. In that case, other uncertainties, such as in the total mass, still dominate.

Prior work has been done on a subset of these fields using an earlier version of the redshift catalog and a different group catalog \citep[e.g.,][]{momcheva06,wong11}.  The formalism we use here is different than used previously; for group halos, we use truncated NFW profiles whereas \citet{momcheva06} use isothermal spheres and \citet{wong11} use untruncated NFW profiles.  \citet{wong11}, \citet{ammons14} and \citet{mccully17} fully model the lens mass distributions \citep[i.e., with GRAVLENS,][]{keeton01} in several of these fields.  

To see how our simple treatment compares, in one field (fbq0951) we calculate $\kappa_{\rm eff}$'s and $\gamma_{\rm eff}$'s for the groups with at least five galaxies and $M_{\rm{vir}} > 10^{13} M_{\odot}$ using the methodology of \citet{ammons14}.  For $\kappa_{\rm eff}$ and $\gamma_{\rm eff} > 0.01$, generally for systems projected closer to the lens and/or with large velocity dispersions, our values agree within the large uncertainties caused by group property uncertainties.  For smaller values of $\kappa_{\rm eff}$ and $\gamma_{\rm eff}$ our values are generally slightly smaller.  This difference can be attributed to us using truncated NFW profiles, which result in smaller halo masses and projected surface mass densities when the halo is projected farther than the truncation radius from the lens sightline rather than the untruncated NFW halos that \citet{ammons14} use.  Our $\kappa_{\rm{tot}}$ value is also slightly lower.  Partly this difference is due to us assuming a different mass profile.  However, it is also partly because we assume that the halos affect the lensing potential independently; we would expect the results from the full treatment, which includes nonlinear effects of interactions between LOS halos \citep{mccully14}, to be different.  Overall, though, the agreement is reasonable; we calculate, using the subset of groups for this field,  $\kappa_{\rm{tot}} =$ 0.23$^{+0.10}_{-0.06}$ and 0.29 and $\gamma_{\rm{tot}} =$ 0.04$^{+0.04}_{-0.03}$ and 0.04 for our formalism and that of \citet{ammons14}, respectively.

\section{Comparison to the literature}
\label{ssec:litcompare}

Several of our systems have been previously identified as needing external shears to explain the image morphology and produce adequately-fitting models.  We compare our calculated shears for these systems (b1422, mg1654, pg1115, and rxj1131) with the values reported for these systems by, respectively, \citet{chiba02} and \citet{nierenberg14} (both for b1422), \citet{wallington96}, \citet{chiba02}, and \citet{claeskens06} and \citet{suyu13} (both for rxj1131).  Our $\gamma_{\rm{tot}}$ values agree with those in the literature within their 3$\sigma$ uncertainties except for field rxj1131.

\citet{claeskens06} find that their best fitting models of rxj1131 include external shear.  Their best model (using a singular isothermal ellipsoid plus an octupole for the lens and external shear) has $\gamma =$ 0.124, much larger than what we calculate ($\gamma_{\rm{tot}} = 0.019^{+0.012}_{-0.010}$).  \citet{suyu13} model an external shear of 0.089$^{+0.006}_{-0.006}$, which is also larger than we measure. However, they find a probability distribution function for the external convergence that peaks around 0.08, in qualitative agreement with our $\kappa_{\rm{tot}} = 0.061^{+0.024}_{-0.016}$. 

The $z \sim$ 0.1 cluster and the lens group both have possible X-ray detections in the literature \citep[2-3 $\times$ 10$^{43}$ ergs s$^{-1}$,][]{morgan06}.  The velocity dispersions we measure are consistent with the observed X-ray luminosity-velocity dispersion relation, which has significant scatter, or larger than what would be predicted given the relation of \citet{ortizgill04}.  So, we are unlikely to be underestimating the lensing contributions due to these two groups.

We only consider lensing due to groups.  Thus, we are not sensitive to additional shear caused by substructures within the lens.  This insensitivity might be why our $\gamma_{\rm{tot}}$ for rxj1131 is significantly lower than that modeled by \citet{claeskens06} and \citet{suyu13}.   \citet{suyu13} downplay the possibility of substructure contributing to their modeled shear based on a modeled external convergence gradient and shear position angle.  However, \citet{keeton09} suggest the presence of substructure because of the order of image arrival time observed in time delay measurements. \citet{cyracine16} show, by modeling dark matter subhalos in lenses, that substructures could have an effect, but it is unlikely to be at the $\gamma \sim 0.1$ level.  Others have suggested the possibility of lens substructure for b1422 and pg1115 \citep[respectively,][]{brada02,miranda07}.  The large uncertainties in our values might wash out the disagreement in these cases.  Dark matter substructures are not the only possible explanation, however.  Baryonic structures, such as disks, which we also do not model, can have similar effects \citep[e.g.,][]{hsueh16,gilman17,hsueh17}.

In addition, extensive lensing analyses on he0435 are being performed by the H0LiCOW team \citep{suyu17}.  \citet{sluse16} conclude using flexion shifts that the groups they identify in this field can be approximated with external shear rather than being modeled explicitly.  \citet{rusu17} constrain $\kappa_{ext}$ to be near zero ($\kappa_{ext}^{med} =$ 0.004), which is also consistent with the $\kappa_{ext}$ determined from an independent weak lensing analysis of this field (Tihhonova et al., in preparation).

Some of this disagreement between their results and ours can be ascribed to differences in group catalogs and lensing methodology.  \citet{wilson16} identify a larger group at $z \sim$ 0.18 than do \citet{sluse16}. Also, \citet{wilson16} use an unweighted centroid, which lies fairly close to the lens and leads to a large convergence $\kappa_{lens} = 0.091^{+0.110}_{-0.031}$. By contrast, \citet{sluse16} use a luminosity weighted centroid, which lies farther from the lens; if we use their centroid, we calculate $\kappa_{lens} =$ 0.035.  These values are still larger than the other estimates of convergence, so \citet{sluse16} suggest that the group centroid may actually lie even farther from the lens sightline, or that the lens is actually at the center of its group halo.  Additional observations of this field might further constrain the group properties and reconcile these differences.

\section{Importance of Lens Groups to External Shear and Convergence}
\label{sec:lensenvirons}

\begin{figure*}
\includegraphics[width=18cm]{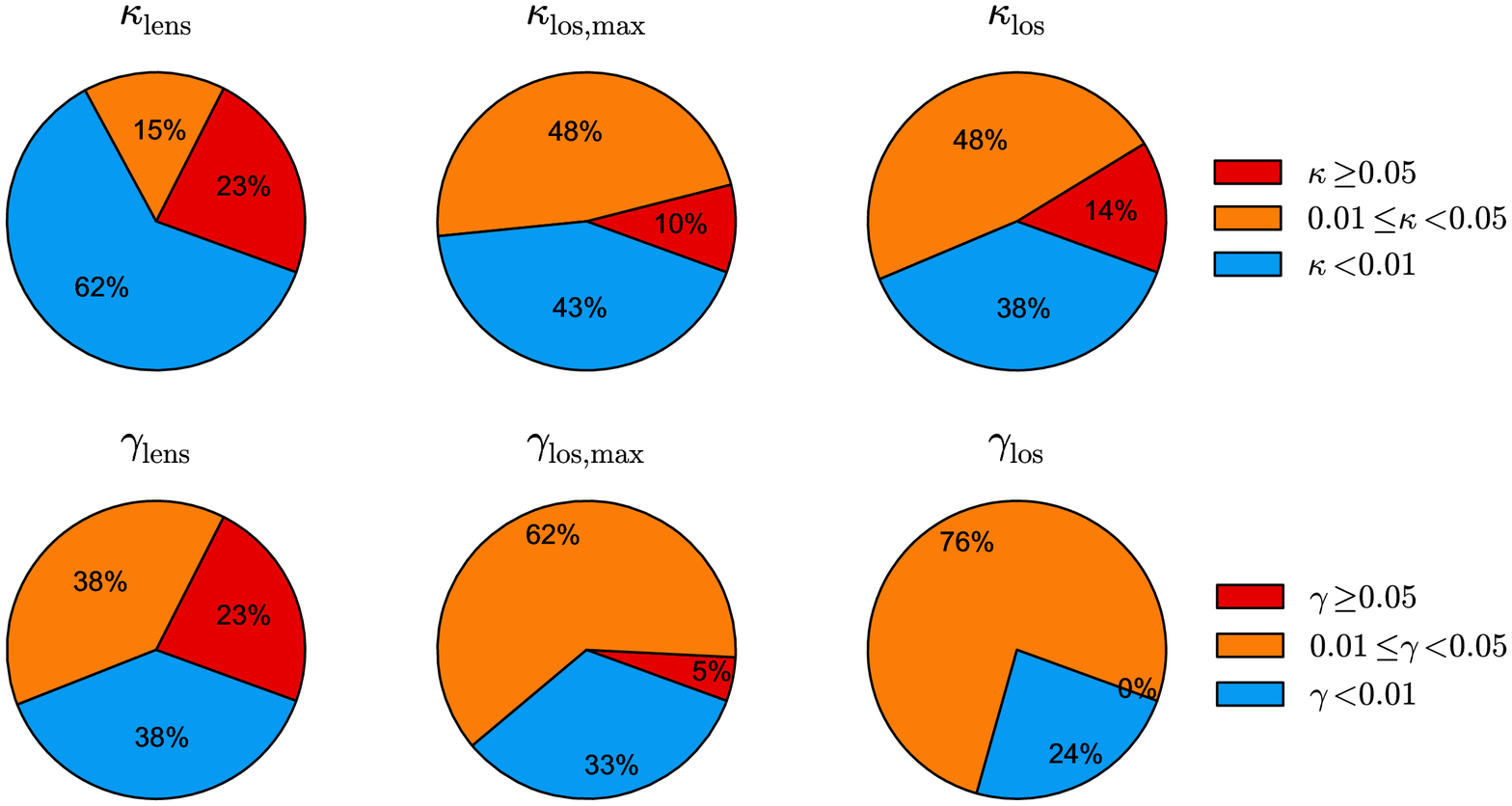}
\caption{The fraction of fields with significant $\kappa$ \textit{(top)} or $\gamma$ \textit{(bottom)} for the lens group (out of 13 fields, \textit{left column}), for the LOS group with the largest contribution (out of 21 fields, \textit{middle}), and for the total LOS excluding the lens group if present (out of 21 fields, \textit{right}). 
}
\label{fig:piecharts}
\end{figure*}

Groups at the lens can have large impacts on the lensing, as they are projected quite close on the sky and are, by definition, very near the lens redshift (presumably any difference being due to peculiar motion).  Thus, we consider here the frequency of important lens groups in our sample.  These results are shown in Figures \ref{fig:fracofkappatot}, \ref{fig:fracofkappatot2}, and \ref{fig:piecharts}.

\subsection{Convergence}
\label{ssec:kappalensenv}

First, we consider $\kappa_{\rm eff}$ and its uncertainty, calculated as described in Section \ref{sec:formalism}, for only lens groups ($\kappa_{\rm{lens}}$, see Table \ref{table:totlensmax}).  Five of 13 lens groups have $\kappa_{\rm{lens}}$ at or above 0.01 and are inconsistent with $<$ 0.01 within their 1$\sigma$ uncertainties.  Three of 13 have $\kappa_{\rm{lens}} \ge$ 0.05.

\begin{figure}
\includegraphics[clip=true, width=9cm]{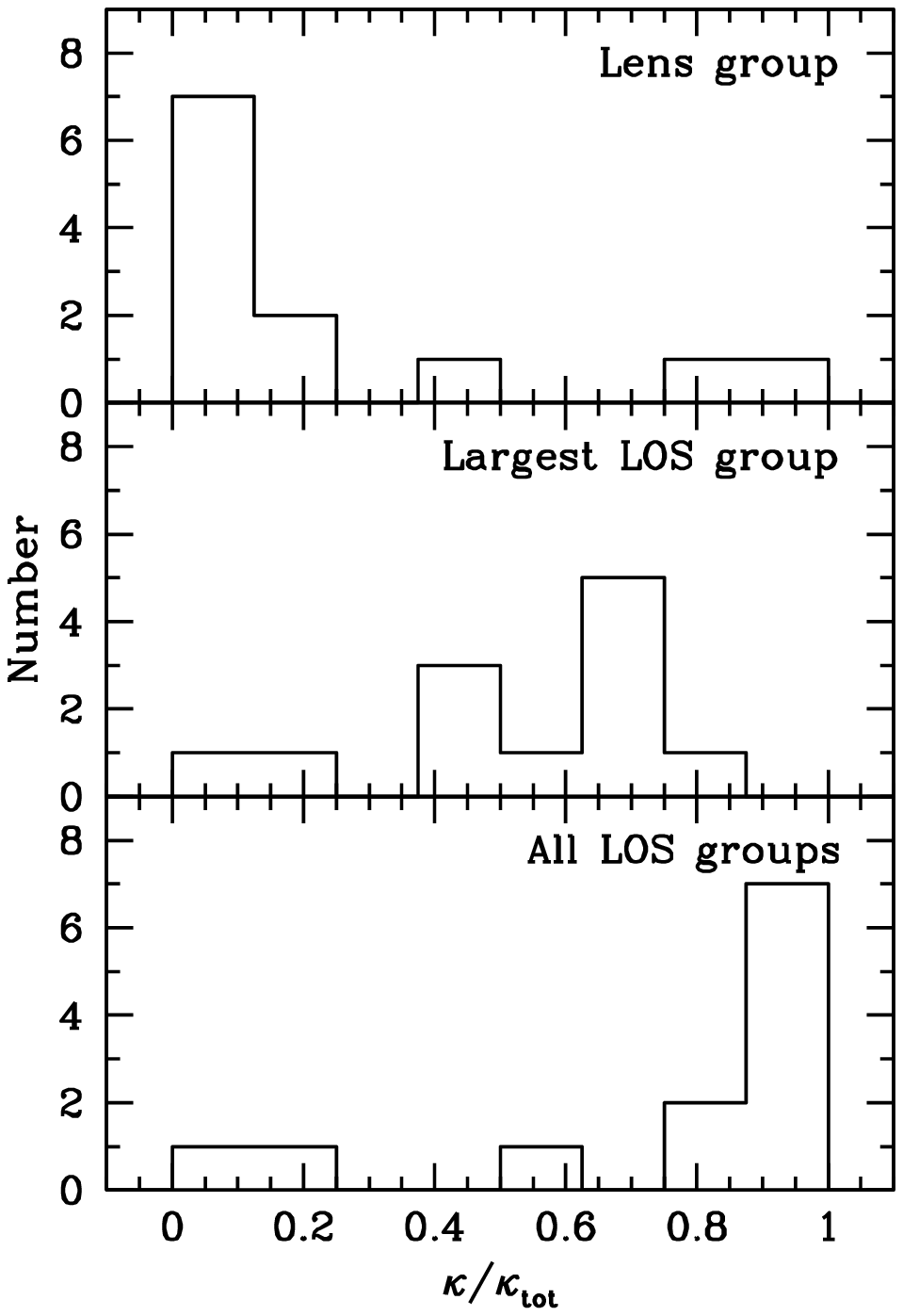}
\caption{The fraction of each field's $\kappa_{\rm{tot}}$ that is contributed by the lens group \textit{(top)}, the LOS group with the largest $\kappa_{\rm eff}$ \textit{(middle)}, and all LOS groups (including the LOS group with the largest $\kappa_{\rm eff}$ but not including the lens group, \textit{bottom}) for fields with $\kappa_{\rm{tot}} \ge$ 0.01 and lens groups. We compare these distributions to test the relative importance of lens groups and the single most important LOS group in fields with both.
 MWW tests comparing the $\kappa_{\rm{lens}}/\kappa_{\rm{tot}}$ distribution with, separately, the $\kappa_{\rm{los,max}}/\kappa_{\rm{tot}}$ and $\kappa_{\rm{los}}/\kappa_{\rm{tot}}$ distributions are significant.  These results suggest that even when lens groups exist they often are not the most important group in the field.
}
\label{fig:fracofkappatot2}
\end{figure}

Next, we consider the fraction of the $\kappa_{\rm{tot}}$ that the lens group contributes (see Figures \ref{fig:fracofkappatot} and \ref{fig:fracofkappatot2}).   While some lens groups dominate their fields, most make up a small fraction of the total convergence once LOS groups are considered; only in two fields do the lens groups make up $>$ 50\% of $\kappa_{\rm{tot}}$.

In summary, several of our fields have significant lens groups, but lens groups rarely are the only important group in the field.

\begin{figure*}
\includegraphics[clip=true, width=13.5cm]{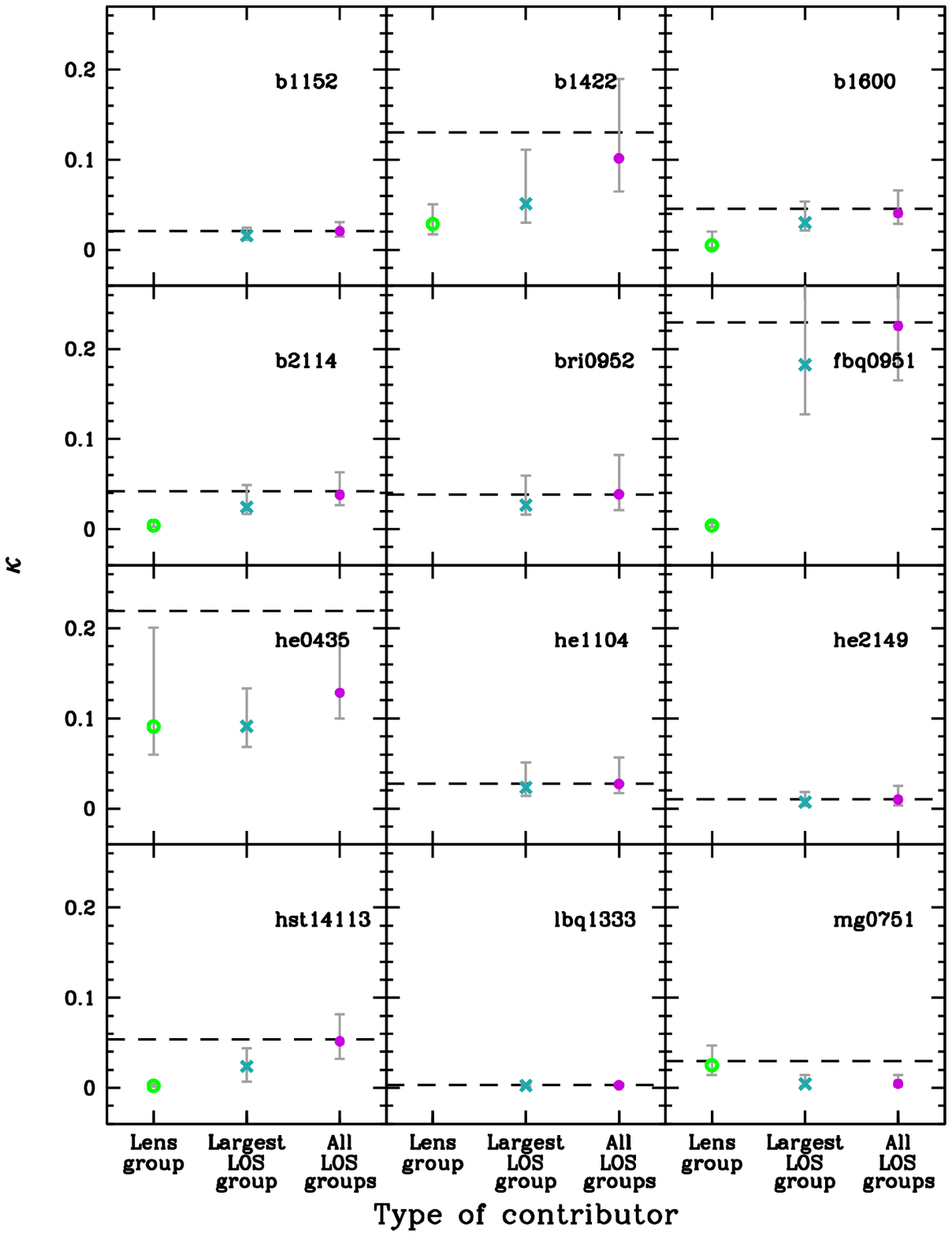}
\caption{The portion of each field's $\kappa_{\rm{tot}}$ that is contributed by lens groups (green open circles), the LOS group with the largest $\kappa_{\rm eff}$ (cyan crosses), and all LOS groups (including the LOS group with the largest $\kappa_{\rm eff}$ but not including the lens group; purple filled circles).  Error bars are calculated from the measurement errors on the group properties using the bootstrap method.  Fields without lens groups have no lens group value shown.  The $\kappa_{\rm{tot}}$ (the sum of the lens and all LOS contributions) is marked with a dashed line.  Although lens groups sometimes are significant, usually LOS groups are important contributors to $\kappa_{\rm{tot}}$.
}
\label{fig:fracofkappatot}
\end{figure*}
\begin{figure*}
\ContinuedFloat
\includegraphics[clip=true, width=13.5cm]{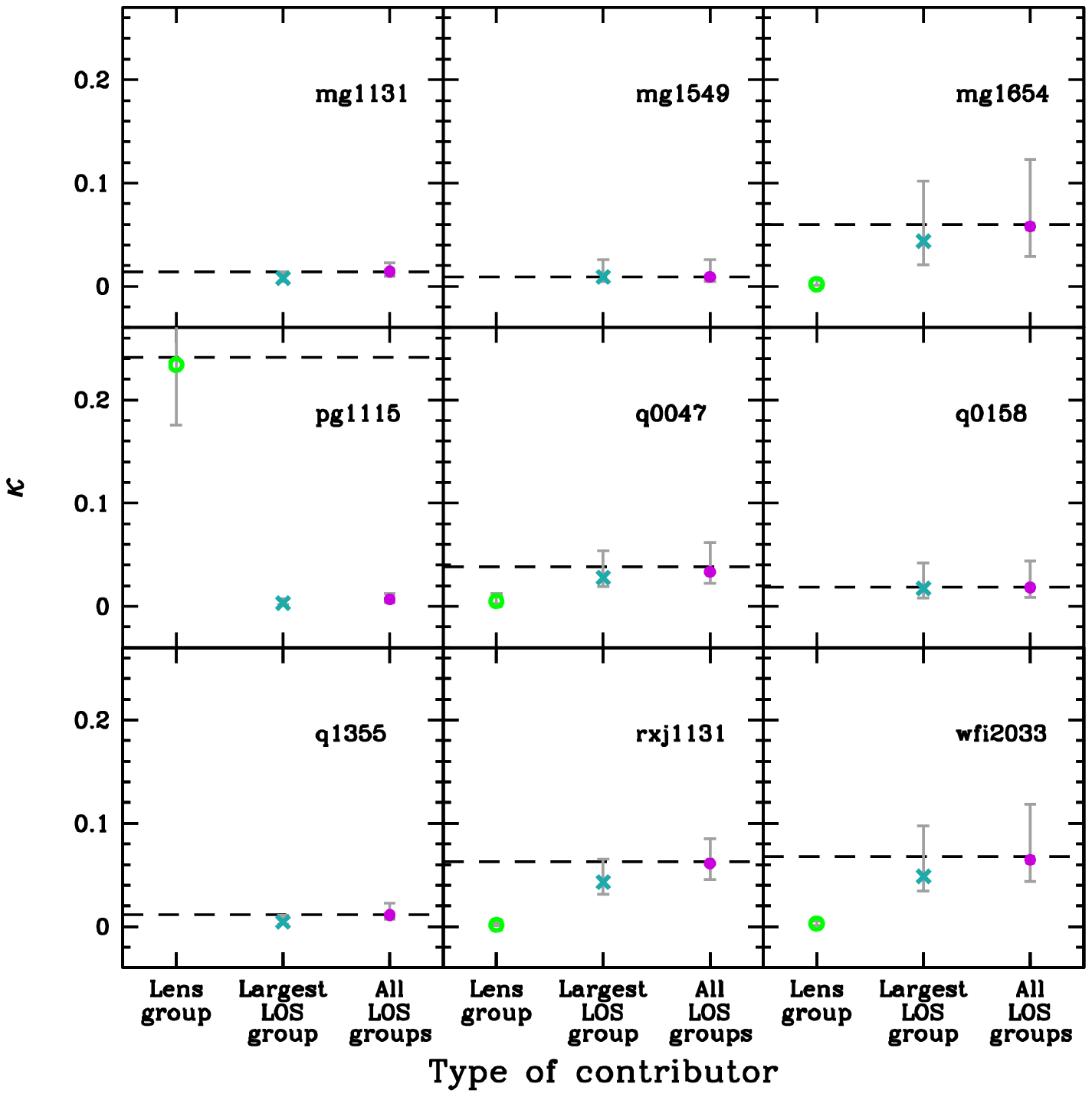}
\caption{Continued.
}
\end{figure*}

\subsection{Shear}

We calculate $\gamma_{\rm eff}$ and its uncertainty for lens groups ($\gamma_{\rm{lens}}$).  Eight of 13 lens groups have $\gamma_{\rm{lens}}$ above 0.01 and are inconsistent with $<$ 0.01 within their 1$\sigma$ uncertainties.  Three of 13 have $\gamma_{\rm{lens}} \ge$ 0.05.

So, as for convergence, several fields have significant shear from lens groups.

\subsection{Lens Environments of Quads Versus Doubles}
\label{ssec:quadvsdoublesenv}

The measured quad to double ratio is higher than expected \citep[e.g.,][]{kochanek96}. Previous studies have suggested that the lens image morphology (i.e., whether there are two or four images) depends on perturbing masses.  \citet{keeton04} find the quad/double ratio is underestimated when lens galaxy environment is neglected.  \citet{momcheva06} also find a suggestion for a link between lens environment and number of images.

The high quad to double ratio also might arise from the lens halos being different shapes.  For example, \citet{collett16} find that quad lenses are more elliptical in their synthetic lens sample.  Yet the ellipticities needed to produce the observed ratio are larger than typical for elliptical galaxies \citep[$b/a =$ 0.4,][]{kochanek96}.  \citet{keeton04} show that both a large lens ellipticity ($e =$ 0.4) and proper consideration of the lens environment are required to reproduce the observed ratio.  Along the same lines, \citet{huterer05} argue that while increasing lens ellipticity or shear increases the quad to double ratio, an additional contribution is needed, most likely from the lens galaxy environment.

No direct selection for quads in groups and/or for doubles not in groups was made when the overall lens sample was constructed.

When we look at double and quad image morphologies here, we do not include any of the systems with ring components.  We do include b2114, which has two image pairs, as a double.  

All seven of our quad lenses with firm spectroscopic redshifts are in groups (including b0712, which we discard elsewhere due to the supergroup), but only three of 10 doubles are in groups.

Using the binomial distribution, we calculate the probability (P-value) of finding at least six of seven quad lenses in groups and no more than three of 10 double lenses in groups, under the hypothesis that the probability of being in a group is the same for both types of lenses.  (We discard the b0712 lower confidence lens group here to be conservative; see Section \ref{sec:grpalg})  The P-value is maximized if the probability of a lens being in a group is 9/17 (which corresponds to the group fraction our combined quad and double sample).  That maximal P-value is 1\%, so this analysis indicates that the hypothesis that quads and doubles have a universal probability of residing in a group can be ruled out at greater than 95\% confidence.

Groups are easier to find if the spectroscopic completeness is higher, and galaxies in the fields of quad systems were given higher priority during spectroscopic followup.  We thus investigate whether this difference in the fraction of lenses in groups is real or an observational bias.  For each field, we calculate the fraction of galaxies in our $I$-band photometry brighter than 20.5 mag, the brighter of the two magnitude limits in our spectroscopic sample, within 5$^{\prime}$ of the lens, the region that was prioritized during spectroscopic followup. We then compare the distributions of spectroscopic completenesses for our quad and double fields.  Neither an F- nor a t-test is significant at the $\ge$ 95\% level.  Thus, there is no evidence for the difference in the frequency of quad and double lenses residing in groups in our sample arising from differences in spectroscopic completenesses in their respective fields.  One of the high redshift lenses (bri0952) is located at a redshift between two groups but is not a member.  We repeat our calculation only using lenses at $z_{lens} <$ 0.6 and the one lens at higher redshift with no evidence of being a group member.  We calculate the probability of finding at least five of six quads lenses in groups and no more than three of eight double lenses in groups if all lenses reside in groups the same fraction of the time to be 4\%.  This result still indicates that quad and double lenses reside in different environments at better than the 2$\sigma$ significance level. 

In addition, we perform a Bayesian analysis to compare two models for our systems at all redshifts: in model M1, the probability of being in a group is the same (f) for quads and doubles (as above); in model M2, the probability of being in a group can be different for quads and doubles (fQ and fD, respectively). We write the Bayesian likelihood as the joint probability of finding 6/7 quads and 3/10 doubles in groups, using the binomial distribution for each sample. For M1, integrating over f with uniform priors yields Bayesian evidence EM1 = 0.0019. For M2, integrating over both fQ and fD with uniform priors yields evidence EM2 = 0.011. According to the \citet{jeffreys61} scale, the Bayes factor EM2/EM1 = 5.9 provides substantial evidence favoring model M2. A similar conclusion is reached using the scale from \citet{kass95}.

Therefore, our quad lens galaxies with firm spectroscopic lens redshifts are more likely to be members of groups than those of doubles; lens environment is correlated with image morphology in our sample.  This result supports the hypothesis that lens environment is important for explaining the larger than expected quad to double ratio.

\section{Importance of Lens LOS Structures to External Shear and Convergence}
\label{sec:losstructureeffects}

Any mass along the LOS will bend light from a background source, so we consider the possible contribution to the lensing potential of groups along our lens lines-of-sight.   These results also are shown in Figures \ref{fig:fracofkappatot}, \ref{fig:fracofkappatot2}, \ref{fig:piecharts}, and \ref{fig:kappaeffquaddoub}.

In future large surveys, scant resources may be available for the deep surveys that would be necessary to create a full group catalog like we use here.  Photometric redshifts of sufficient quality to identify peaks in the LOS redshift distribution that indicate possible clusters and large groups likely will be available, however.  So, we investigate whether prioritizing spectroscopic followup of the most significant LOS group would capture most of the LOS contribution from groups. 

Ideally, a lens field would have extensive spectroscopic followup which could be used to identify and characterize all the LOS groups rather than just the most significant one.  So, we also investigate the full LOS group distribution.  We test how significant the LOS groups are as well as look for fields with a significant total LOS contribution but without any individually significant groups.

Although groups with larger convergences typically also have larger shears, there is some scatter in the relation.  Thus, the group with the largest convergence is not necessarily that with the largest shear.  In only one field of our 21 field subset is the LOS group with the largest $\kappa_{\rm eff}$ not also the LOS group with the largest $\gamma_{\rm eff}$. 

\subsection{Convergence}
\label{ssec:kappalos}

We look for LOS groups with the largest $\kappa_{\rm eff}$ in their fields that are individually significant ($\kappa_{\rm{los,max}}$ groups, see Table \ref{table:totlensmax}).  There are 12 fields with $\kappa_{\rm{los,max}} \ge$ 0.01 including the 1$\sigma$ uncertainty.  Two fields have $\kappa_{\rm{los,max}} \ge$ 0.05.

We then calculate the contribution of $\kappa_{\rm{los,max}}$ groups to $\kappa_{\rm{tot}}$ for fields with $\kappa_{\rm{tot}} \ge$ 0.01 (see Figures \ref{fig:fracofkappatot} and \ref{fig:fracofkappatot2}).  As was the case for lens groups (see Section \ref{ssec:kappalensenv}), the $\kappa_{\rm{los,max}}$ groups span a large range of contributions to $\kappa_{\rm{tot}}$.  However, in fields with lenses the lens groups' distribution has a mean value of 25\% of the total, while that for the $\kappa_{\rm{los,max}}$ groups has a mean of 53\%; an MWW test is significant.  In eight of 12 fields with lens groups (67\%), the fraction of $\kappa_{\rm{tot}}$ contributed by the $\kappa_{\rm{los,max}}$ group is greater than two times that contributed by the lens group.  We perform a bootstrap means comparison (see Appendix \ref{appendix:bootmean}).  The resulting mean for the lens groups' fraction of $\kappa_{\rm{tot}}$ distribution is as large or larger than that for the $\kappa_{\rm{los,max}}$ groups in 0.3\% of trials.  The mean of the $\kappa_{\rm{los,max}}$ groups fraction of $\kappa_{\rm{tot}}$ distribution is at least as small as the measured value for lens groups $<$ 0.1\% of the time.  So, the means of these distributions are significantly different.

Thus, lens groups, when present, often are not the most important single group in their fields.  This result agrees with our finding in Section \ref{ssec:kappalensenv} that lens groups often are not the only important group-scale contributor. 

We compare the properties of lens groups and the $\kappa_{\rm{los,max}}$ groups.   Eight out of 12 (67\%) of the $\kappa_{\rm{los,max}}$ groups are in the foreground of the lenses.  The $\kappa_{\rm{los,max}}$ groups and lens groups do not have mean redshifts distinguishable at $\ge$ 95\% confidence level in a t-test, however.  This result indicates the $\kappa_{\rm{los,max}}$ groups are not at significantly lower redshifts where they might be better sampled in our data. The mean $\kappa_{\rm{los,max}}$ group velocity dispersion is significantly larger than those of lens groups using a t-test.  Although galaxy clusters are uncommon, these sightlines probe a large volume.  Clusters are also more common at lower redshifts where we have few lens galaxies.

Next, we calculate the convergence due to all LOS groups ($\kappa_{\rm{los}}$).  Thirteen fields are consistent with $\kappa_{\rm{los}} \ge$ 0.01, and three fields are consistent with $\kappa_{\rm{los}} \ge$ 0.05, considering the 1$\sigma$ uncertainties.  The fractional contribution of $\kappa_{\rm{los}}$ to $\kappa_{\rm{tot}}$ is shown in Figure \ref{fig:fracofkappatot} and, for just fields with lens groups, Figure \ref{fig:fracofkappatot2}.  The mean of the $\kappa_{\rm{los}}/\kappa_{\rm{tot}}$ distribution for fields with lens groups  (75\%) is larger than that of $\kappa_{\rm{lens}}/\kappa_{\rm{tot}}$  (25\%).  The distributions are distinguishable using an MWW test.  Again, we perform a bootstrap means comparison.  The resulting mean for the lens groups' fraction of $\kappa_{\rm{tot}}$ distribution is as large or larger than that for the $\kappa_{\rm{los}}$ groups $<$ 0.1\% of the time, and the mean of the $\kappa_{\rm{los}}$ fraction of $\kappa_{\rm{tot}}$ distribution is at least as small as the measured value for lens groups $<$ 0.1\% of the time.  So, the means of these distributions also are significantly different.

All of our fields have multiple groups identified.  If a field has enough groups, fields without any that are significant by themselves could nonetheless have a significant $\kappa_{\rm{tot}}$ due to the combination of multiple individually insignificant groups.  Of those fields without either a lens or LOS group with $\kappa_{\rm eff} \ge$ 0.01 considering the 1$\sigma$ uncertainties, one has a $\kappa_{\rm{tot}} \ge$ 0.01.  One field has $\kappa_{\rm{tot}} \ge$ 0.05, despite not having any individual groups meeting this criterion.  These results show it is possible for multiple insignificant groups to add up to a significant total contribution, although the dominant mode is for a significant overall contribution to be driven by at least one individually significant group.  Our analysis throughout this paper only provides a lower limit to how frequently the former mode occurs, however (see Section \ref{sec:formalism}).

\subsection{Shear}

There are 14 fields where the LOS group with largest $\gamma_{\rm eff}$ ($\gamma_{\rm{los,max}}$) is consistent with being greater than 0.01 considering the 1$\sigma$ uncertainties (see Table \ref{table:totlensmax}).  One field has $\gamma_{\rm eff}$ greater than 0.05.

Again, we look at the shear contribution due to all LOS groups ($\gamma_{\rm{los}}$).  Sixteen fields have $\gamma_{\rm{los}} \ge$ 0.01, and none have $\gamma_{\rm{los}} \ge$ 0.05, considering the 1$\sigma$ uncertainties.

As for the convergence, we look for fields without an individually significant group that nevertheless has a significant total shear.  Of those groups without either a lens or LOS group with $\gamma_{\rm eff} \ge$ 0.01 considering the 1$\sigma$ uncertainties, four have $\gamma_{\rm{tot}} \ge$ 0.01.  One field has $\gamma_{\rm{tot}} \ge$ 0.05, considering the uncertainties, despite not having any individual groups meeting this criterion.  These results support those for the convergence that sometimes, but not often, a field can have an overall shear due to groups that is significant without having any individually significant groups.

\subsection{LOS Structures of Quads Versus Doubles}
\label{ssec:quadvsdoubleslos}

Since LOS structures are important to our fields as a whole, we now test whether these structures affect the quad to double ratio, as the lens environments did in Section \ref{ssec:quadvsdoublesenv}.

\begin{figure*}
\includegraphics[clip=true, width=18cm]{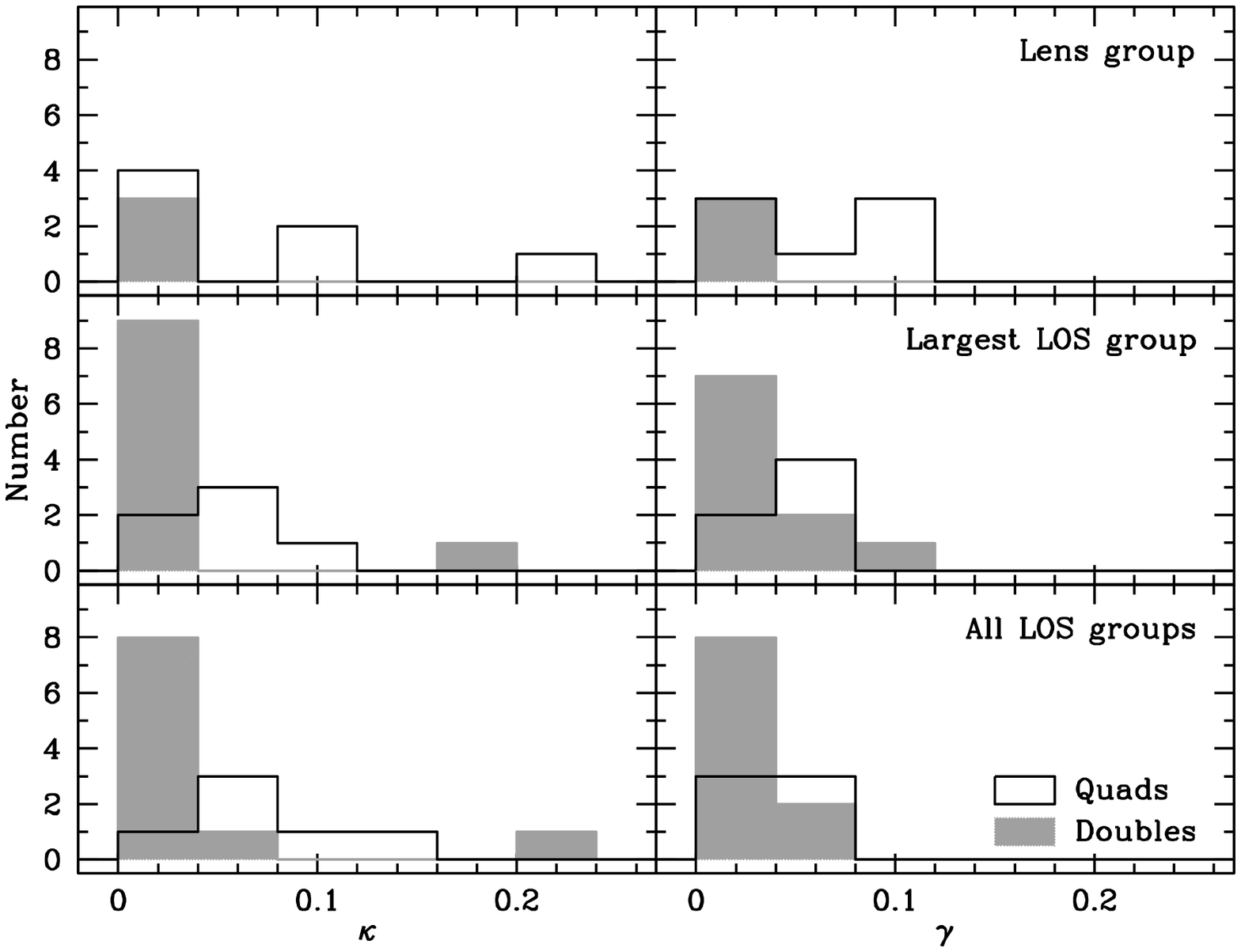}
\caption{Distributions of $\kappa$ (\textit{left}) and $\gamma$ (\textit{right}) for fields with quads (white histograms) versus those with doubles (solid gray).  \textit{Top:} $\kappa_{\rm eff}$ and $\gamma_{\rm eff}$ for lens groups only.  The b0712 lens group is included in this row, but not in the LOS rows below due to the supergroup in this field.  The highest convergences arising from lens groups are in quads.   \textit{Middle:} $\kappa_{\rm eff}$ and $\gamma_{\rm eff}$ for LOS groups with the largest of these values in their field.  The quad versus double distributions cannot be distinguished.  \textit{Bottom:} $\kappa_{\rm los}$ and $\gamma_{\rm los}$ distributions or all groups in the sightline except for the lens group.  The shapes of the $\kappa_{\rm los}$ distributions are different for quads versus doubles, but the means are not.  In summary, while lens environment is connected to the observed high quad to double ratio, we currently have little evidence for the importance of LOS groups in this context.
}
\label{fig:kappaeffquaddoub}
\end{figure*}

First, we consider the quad versus double distributions of $\kappa_{\rm{los}}$, which includes all structures along the sightline except for at the lens (left bottom panel; Figure \ref{fig:kappaeffquaddoub}). A K-S test excludes the possibility that the quad and double distributions are drawn from the same distribution at the $>$95\% level.  However, a bootstrap means comparison test (see Appendix \ref{appendix:bootmean}) does not distinguish the means.

Next, we compare the LOS groups with the largest $\kappa_{\rm eff}$ (left middle panel; Figure \ref{fig:kappaeffquaddoub}) in the quads versus double fields.  The $\kappa_{\rm{los,max}}$ distributions of these groups in the fields of quads and doubles are not distinguishable using a K-S test, nor are their means using a bootstrap means comparison.

For completeness, we also consider the shear distributions for the LOS groups with the largest $\gamma_{\rm eff}$ ($\gamma_{\rm{los,max}}$), and the full LOS (right middle and right bottom panels; Figure \ref{fig:kappaeffquaddoub}).  Neither the $\gamma_{\rm{los}}$ nor the $\gamma_{\rm{los,max}}$ quad versus double distributions are distinguishable using K-S test or a bootstrap means comparison.

These results agrees with those of \citet{collett16}, who find a only a small bias (a 0.009 shift in the LOS convergence expectation value) in LOS properties of quad sightlines when the lens environment is excluded.

In summary, while some LOS groups can be more important to $\kappa_{\rm{tot}}$ than the lens group (Section \ref{ssec:kappalos}), quad lenses are more often in groups than doubles (Section \ref{ssec:quadvsdoublesenv}) and those quad groups can produce larger $\kappa_{\rm eff}$'s (Figure \ref{fig:kappaeffquaddoub}).  These two results together suggest that the local lens environment plays a significant role in driving the quad to double ratio.  Any additional role of the LOS in increasing the quad to double ratio remains unclear.

\section{Implications for $H_0$}
\label{sec:H0}

To summarize, at least five of 22 (23\%) of our systems with firm spectroscopic lens redshifts (including b0712) have lens groups with effective convergences that affect the lensing potential at the $\ge$ 1\% level or greater \citep[assuming the lens galaxy's convergence is $\sim$ 1; see, e.g.,][]{keeton04}.  Twelve of 21 (57\%) of our fields with firm lens spectroscopic redshifts and no supergroup have at least one LOS group not at the lens redshift that contributes at the $\ge$ 1\% level.  The dominant contributor to $\kappa_{\rm{tot}}$ for fields with both a significant $\kappa_{\rm{tot}}$ and a lens group is more frequently a LOS group than a lens group.  Considering all groups, 15 of 21 (71\%) fields have $\kappa_{\rm{tot}}$ that will affect time-delay lens measurements of $H_0$ at the $\ge$ 1\% level.

Often modelers assume that the lens galaxy and the lens environment are the only two important contributors to the lensing potential.  Our results indicate that while the lens environment can be important, in a substantial fraction of systems there also is significant LOS structure.  Thus, if either lens groups or LOS groups are neglected, there could be significant systematic bias \citep[an overestimate, since $h \propto (1-\kappa)$, see][]{keeton04} in the derived $H_0$ if 1\% accuracy is the goal.  

As it will be observationally expensive to perform deep spectroscopic surveys in all of the several thousand lens fields LSST is expected to find, an alternative method to account for group-scale perturbations would be to statistically correct the derived $H_0$ given some average convergence.  This general approach of calibrating a lensing parameter is already being used; \citet[][]{rusu17} use sightlines through simulations that are similar to observed lens sightlines to constrain the external convergence of individual lens systems, for example.  Here, we calculate the average convergences of the nine time delay lens systems in our sample.  If the convergences from lens environments and LOS groups are ignored, the resulting $H_0$ would be overestimated by 11$^{+3}_{-2}$\% on average.  If only LOS groups are ignored, then $H_0$ would be overestimated by 7$^{+3}_{-2}$\%.  This calibration will improve with larger and thus more representative samples. 

\section{Calibrating the $\kappa_{tot}$ - $\gamma_{tot}$ Relation}
\label{sec:gammatotvkappatot}

Although $\gamma_{\rm{tot}}$ can be inferred from lens models of measured image separations, $\kappa_{\rm{tot}}$ is unconstrained because of the mass sheet degeneracy. However, neglecting the contributions of lens environment and LOS structures to $\kappa_{\rm{tot}}$ can bias derived parameters, such as $H_0$ (as discussed in Section \ref{sec:H0}) and $\Omega_{\Lambda}$ \citep{keeton04}.  Here we compare the measured $\kappa_{\rm{tot}}$ to the measured $\gamma_{\rm{tot}}$ from our fields to quantify their relationship and to estimate how predictive a lens-model derived $\gamma_{\rm{tot}}$ might be of that system's $\kappa_{\rm{tot}}$.

We calculate both $\kappa_{\rm{tot}}$ and $\gamma_{\rm{tot}}$ for each of our lens fields using our group catalog.  We note that our dynamical measure of $\gamma_{\rm{tot}}$ can differ from that derived via lens modeling of image separations \citep[see][]{wong11}.  Our shear is by definition large-scale.  If there is any small-scale shear due to the structure of the main lens galaxy itself (e.g., a misalignment of the stellar and dark matter components, dark matter substructure, or baryonic structures) it would be reflected in the lens modeling derived shear but not our calculation of $\gamma_{\rm{tot}}$.  However, our values agree within the uncertainties for three of the four systems for which lens derived measurements are available in the literature (see Section \ref{ssec:litcompare}).

\begin{figure}
\includegraphics[clip=true, trim= 9.5cm 9.5cm 0cm 0cm, width=9cm]{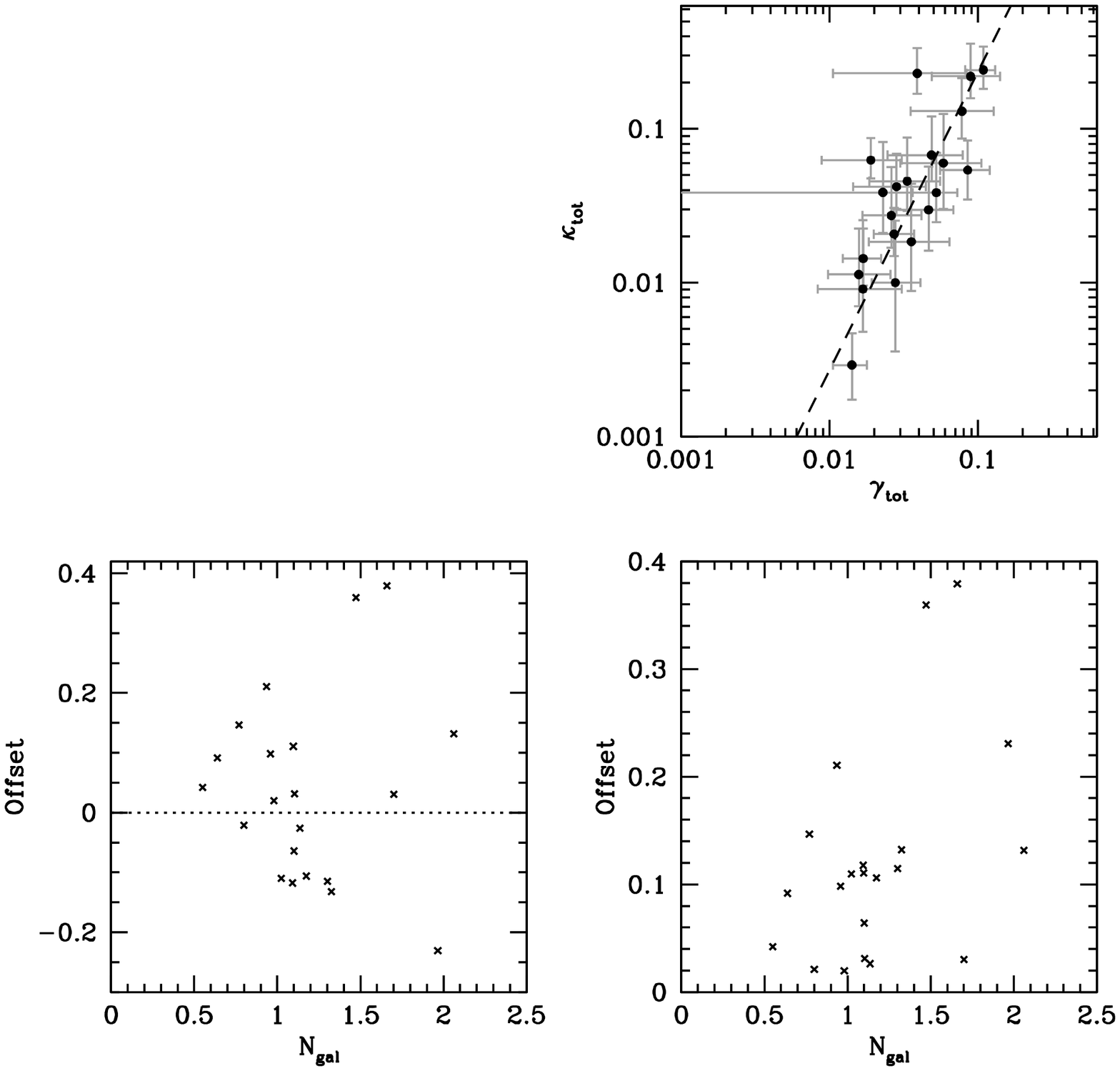}
\caption{$\kappa_{\rm{tot}}$ versus $\gamma_{\rm{tot}}$ for our 21 field subsample.  Error bars denote the 16th and 84th percentiles on our 1000 bootstrap iterations of our group catalog.  The black dashed line ($\log(\kappa_{\rm{tot}}) = (1.94 \pm 0.34) \log(\gamma_{\rm{tot}}) + (1.31 \pm 0.49)$) is a $\chi^2$ fit including the geometric means of the two-sided error bars in both $\kappa_{\rm{tot}}$ and $\gamma_{\rm{tot}}$.  A small shear does not necessarily imply the field also has a small convergence, but a large shear generally implies a large convergence; four of the 21 fields (19\%) have $\kappa_{\rm{tot}} > 2\gamma_{\rm{tot}}$. 
}
\label{fig:kappatotgammatotscatter}
\end{figure}

There should be a physical relation between shear and convergence, since they both arise from the mass distribution.  Being able to estimate convergence from the shear inferred from lens models, perhaps with some adjustments from photometric observables like galaxy number density \citep[e.g.,][]{suyu13}, could help select ``golden lenses,'' i.e., low convergence systems, from a large lens sample, thus prioritizing the more laborious spectroscopic follow-up.  Here we use our measurements to calibrate directly the relation between $\kappa_{tot}$ and $\gamma_{tot}$.

In Figure \ref{fig:kappatotgammatotscatter} we plot the $\kappa_{\rm{tot}}$ and $\gamma_{\rm{tot}}$ for our fields and fit a power law to the relation.  We find the geometrical mean of the two-sided error bars on $\kappa_{\rm{tot}}$ and $\gamma_{\rm{tot}}$, which were calculated from the uncertainties in measured group properties using the bootstrap method (see Section \ref{sec:formalism}). The $\chi^2$ fit using the resulting uncertainties, and the uncertainties in the slope and intercept, is $\log(\kappa_{\rm{tot}}) = (1.94 \pm 0.34) \log(\gamma_{\rm{tot}}) + (1.31 \pm 0.49)$.  The root mean square error is 0.34 dex.  The linear Pearson correlation coefficient of $\rm{log}(\kappa_{\rm{tot}})$ versus $\rm{log}(\gamma_{\rm{tot}})$ is 0.77.

The slope of the relation is steeper than $\kappa = \gamma$. This slope is affected by the shape of the assumed density profiles of the individual group halos. At small shears and convergences, the $\gamma_{eff}$ of individual halos tends to be larger than $\kappa_{eff}$ for NFW halos.  This relation between the individual halo values approaches $\kappa = \gamma$ then becomes more steep at large convergence and shear because of the shape of the inner regions of NFW halos \citep[see Appendix \ref{appendix:truncationchoice} and][]{oguri05}.  These values are combined differently to calculate $\kappa_{\rm{tot}}$ and $\gamma_{\rm{tot}}$; convergences simply add, but shears add as tensors.  As groups' contributions are added up, convergences always increase, but shears can increase or decrease depending on the groups' relative positions on the sky.  As the strength of the contribution from multiple LOS perturbers increases, fields consequently are pushed to the upper left of the $\kappa = \gamma$ relation.  If a field's large convergence and shear is due to a single structure, then the total field values do not necessarily move above the relation.  Fields with few and/or very insignificant groups may remain below the $\kappa = \gamma$ relation if the sum of the $\kappa_{eff}$'s is not enough to overcome their initial deficit compared to $\gamma_{eff}$.

Recently, investigators have used a combination of galaxy counts and cosmological simulations to constrain $\kappa_{\rm{tot}}$.  For example, \citet{suyu13} calculate the galaxy counts, external shears, and convergences in 64 simulated sightlines in the Millennium Simulation and then select the sightlines that best match the observed lens sightline to constrain the likely value of convergence external to the main galaxy lens. \citet{greene13} and  \citet{rusu17} use a similar technique but with weighted number counts as well as external shear in the latter study.  

Does incorporating the normalized galaxy number density along the sightline to reduce the scatter in our $\kappa_{tot}$ versus $\gamma_{tot}$ relation?  We would expect lens lines-of-sight to have larger overdensities, since most lens galaxies are massive galaxies that preferentially reside in higher-density environments \citep[e.g.,][]{hilbert07,fassnacht11}.  To calculate this overdensity, or normalized galaxy number density ($N_{gal,norm}$), we divide the number of galaxies in our photometry with $I <$ 21.5 mag within 2$^{\prime}$ of the lens by the median number of such galaxies in random sightlines chosen in the following way in the same field.

We examine all the objects, stars as well as galaxies, in the photometric fields to mask out regions that could have depressed galaxy counts, including blank regions from non-operational CCDs, obvious chip gaps due to inadequately dithered exposures, and regions near bright stars.  We mask out the lens sightlines when selecting random sightlines.  We include only random sightlines that lie completely within the well-sampled field.  We discard random sightlines that overlap the masked regions until we have 1000 acceptable sightlines per field. 

We calculate the partial correlation coefficient of log($\kappa_{\rm{tot}}$) versus log($\gamma_{\rm{tot}}$) removing the effects of $N_{gal,norm}$.  The value, 0.77, is the same as without considering $N_{gal,norm}$.  There is also not a significant correlation between the offset of a field from the best-fit line and $N_{gal,norm}$.  Thus, it is not obvious that incorporating the normalized galaxy number density improves our correlation.

Overall, the $\gamma_{\rm{tot}}$ does not always predict $\kappa_{\rm{tot}}$ well.  Four of 21 fields (19\%) have $\kappa_{\rm{tot}}/\gamma_{\rm{tot}} \ge$ 2; all have $\kappa_{\rm{tot}}, \gamma_{\rm{tot}} >$ 0.01.  The $\kappa_{\rm{tot}}$ inferred directly from $\gamma_{\rm{tot}}$ will thus be underestimated by $\ge$ 0.01, leading to a $\ge$ 1\% bias in $H_0$.  Our $\gamma_{\rm{tot}}$ and $\kappa_{\rm{tot}}$ measurements have large uncertainties, so how far a field might actually be off the relation is uncertain.  However, in our bootstrap iterations of our group catalog, 16\% have $\kappa_{\rm{tot}}/\gamma_{\rm{tot}} \ge$ 2, and 0.5\% have $\kappa_{\rm{tot}}/\gamma_{\rm{tot}} \ge$ 10.

If only the lens group contribution to the shear is known, one should be especially cautious of using this shear to predict the field's $\kappa_{\rm{tot}}$.  For our measured groups, 10 of 12 fields with lens groups (83\%) have $\kappa_{\rm{tot}}/\gamma_{\rm{lens}} \ge$ 2.  One of 12 (8\%) has $\kappa_{\rm{tot}}/\gamma_{\rm{lens}} \ge$ 10.  In our bootstrap iterations, 63\% have $\kappa_{\rm{tot}}/\gamma_{\rm{lens}} \ge$ 2, and 17\% have $\kappa_{\rm{tot}}/\gamma_{\rm{lens}} \ge$ 10.  

The shear of the LOS group with the largest shear is better at predicting the field's $\kappa_{\rm{tot}}$.  For our measured groups, 5 of 21 fields (24\%) have $\kappa_{\rm{tot}}/\gamma_{\rm{los,max}} \ge$ 2.  One of 21 (5\%) has $\kappa_{\rm{tot}}/\gamma_{\rm{los,max}} \ge$ 10.  In our bootstrap iterations, 25\% have $\kappa_{\rm{tot}}/\gamma_{\rm{los,max}} \ge$ 2, and 3\% have $\kappa_{\rm{tot}}/\gamma_{\rm{los,max}} \ge$ 10. 

\begin{figure*}
\includegraphics[clip=true,  width=18cm]{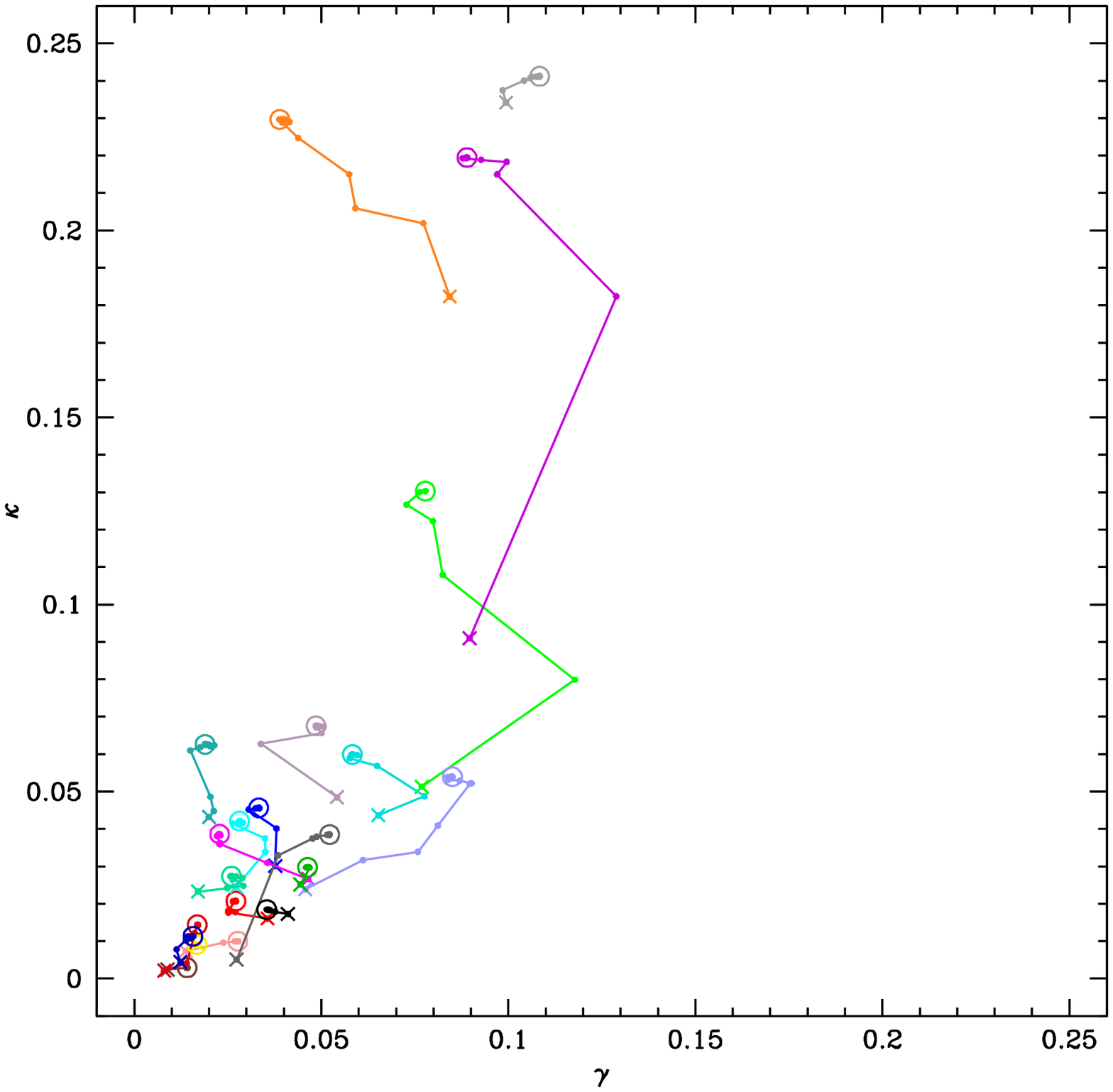}
\caption{Build up of $\gamma_{\rm{tot}}$ and $\kappa_{\rm{tot}}$ from individual groups for each lens beam.  Each colored track represents an individual field.  The cross indicates the $\gamma_{\rm eff}$ and $\kappa_{\rm eff}$ of the group with the largest $\gamma_{\rm eff}$ in that field.  Each connected dot represents the $\gamma_{\rm{tot}}$ and $\kappa_{\rm{tot}}$ each time a group with the next smallest $\gamma_{\rm eff}$ is added.  The $\gamma_{\rm eff}$ and $\kappa_{\rm eff}$ with all groups in the field included is marked with an open circle.  There is a diversity of ways of reaching a field's total values depending on how many comparably significant groups there are and their relative position angles on the sky.
}
\label{fig:stepstokappagammatot}
\end{figure*}

Figure \ref{fig:stepstokappagammatot} illustrates how many groups are significant contributors to $\kappa_{\rm{tot}}$ and $\gamma_{\rm{tot}}$, as well as how adding each group affects $\gamma_{\rm{tot}}$ and $\kappa_{\rm{tot}}$.  We sort the groups within a field from largest $\gamma_{\rm eff}$ to smallest and calculate $\gamma_{\rm{tot}}$ and $\kappa_{\rm{tot}}$ after adding each group one at a time.   We plot the resulting tracks in Figure \ref{fig:stepstokappagammatot}.

There is a diversity of tracks.  As is also evident in Figures \ref{fig:loslensingeffectskappa} and \ref{fig:gammavecfield}, some fields have one group that dominates, others have one or two that are of similar importance, and others have several significant groups.  Figure \ref{fig:gammavecfield} illustrated how the distribution of groups on the sky in our fields can affect the $\gamma_{\rm{tot}}$, as well.  Some fields' groups with the largest $\gamma_{\rm eff}$ have similar $\gamma_{\rm eff}$ and $\kappa_{\rm eff}$ but end up with total field values that are in different parts of the plot (e.g., the magenta, green, and light blue fields in Figure \ref{fig:stepstokappagammatot}).  Other fields have total values that are similar but have groups with the largest $\gamma_{\rm eff}$ with $\gamma_{\rm{los,max}}$ and $\kappa_{\rm{los,max}}$ that were different (e.g., the gray and purple fields near the top of Figure \ref{fig:stepstokappagammatot}).  

From this analysis, the scatter and best-fit relation of log($\gamma_{\rm{tot}}$) and log($\kappa_{\rm{tot}}$) appear to be driven by the number of groups and their relative placement on the sky.  Two fields can have similar $N_{gal,norm}$ but much different $\gamma_{\rm{tot}}$ because the two most important groups are oriented differently on the sky ($\sim$ 90$^{\circ}$ apart around the lens instead of $\sim$ 180$^{\circ}$ or 0$^{\circ}$ apart) or because the galaxies are mostly in one group (which then dominates the lensing perturbation) rather than split into many (which can have canceling $\gamma_{\rm eff}$'s).

\section{Conclusions}

Cosmological constraints on $H_0$ from strong gravitational lensing measurements now aim for percent or even sub-percent precision, a goal directly depending on the uncertainty in the convergence ($\kappa$).  Unlike shear ($\gamma$), $\kappa$ is not directly measured by lensing observations, only inferred from assumptions about the mass distribution.  Here we use kinematic measurements of the mass distribution of the lens environment and along the line-of-sight to constrain $\kappa$ directly, quantifying how much ignoring the contributions of lens environment and line-of-sight structures might bias $H_0$ determinations.  We also determine $\gamma$ from our kinematic measurements, quantifying the effects on $\gamma$ due to lens environment and the line-of-sight and comparing them to the precision of current $\gamma$ measurements from lens imaging.  Since shears enter the lens equation as $\gamma$ multiplied by the uncertainty in lens image position \citep[typically $\sim 0.^{\prime\prime}$003;][]{courbin97}, shears of 0.01 lead to $\sim 3\sigma$ effects.  We also evaluate the usefulness of $\gamma$ as a predictor of $\kappa$.

To test these effects on $\kappa$ and thus on $H_0$, we consider two significance thresholds: $\kappa \ge$ 0.01 and 0.05.  The former is consistent with the percent target on $H_0$ \citep{lsstcolab12}, the latter is the approximate level of current discrepancies among $H_0$ determinations from different methods \citep[e.g.,][]{riess11,bennett14,plank16H0,wong17}.  We consider the same thresholds for assessing significant lens environment or line-of-sight contributions to $\gamma$, given that shears of $\ge$ 0.01 correspond to $\sim 3\sigma$ effects compared with the observed image position uncertainties, making heretofore ignored contributions at the $>$0.01 level problematic.

In \citet{wilson16}, we identified massive groups around the lens galaxy and along the sightline in the fields of 26 strong lensing galaxies.  Here, we assess the significance of those lens' environments and lines-of-sight in our cleanest 21 fields.

Our main results are as follows:

\begin{itemize}

\item
The total convergence due to any group at the lens and along the sightline ranges from $\kappa_{\rm{tot}} <$ 0.01 to 0.24.  The total shear $\gamma_{\rm{tot}}$ ranges from $<$ 0.01 to 0.11.

\item
Lens groups are often important:  five of 22 (23\%) have $\kappa_{\rm eff} \ge$ 0.01, and three (14\%) have $\kappa_{\rm eff}​ \ge$ 0.05.  Considering shear, eight (36\%) have $\gamma_{\rm eff} \ge$ 0.01, and three (14\%) have $\gamma_{\rm eff} \ge$ 0.05.

\item
Line-of-sight groups (not at the lens redshift) are often important as well.  In 12 of 21 fields (57\%), there is at least one line-of-sight group with $\kappa_{\rm eff} > 0.01$, and 2 (10\%) have $\kappa_{\rm eff}​ \ge$ 0.05.  Fourteen fields (67\%) have at least one line-of-sight group with $\gamma_{\rm eff} \ge$ 0.01, and 1 (5\%) has $\gamma_{\rm eff} \ge$ 0.05.

\item 
Line-of-sight groups can be more important than the lens group.  In fields with lens groups and $\kappa_{\rm{tot}} \ge$ 0.01, the line-of-sight group with the largest $\kappa_{\rm eff}$ contributes $\ge$ 2$\times$ more to $\kappa_{\rm{tot}}$ than the lens group $\sim$ 67\% of the time.

\item
Previous studies \citep[e.g.,][]{keeton04,momcheva06} suggest that lens environment is connected to the relative numbers of systems with four and two images (the quad/double ratio) and why more quads are observed than expected \citep{kochanek96}.  All seven of our quad lenses, but only three of 10 doubles, are in groups, a  statistically significant difference.  There is a population of quad lens groups with large $\kappa_{\rm eff}$ that has no counterpart for doubles.  Thus, lens groups are important for the quad to double ratio.  The effects of the line-of-sight structures are not as obvious: we do not find a statistically significant difference between the line-of-sight convergences for quads versus doubles.

\item

Without proper calibration, shear can be a poor predictor of convergence.  The correlation is significant with the form $\log(\kappa_{\rm{tot}}) = (1.94 \pm 0.34) \log(\gamma_{\rm{tot}}) + (1.31 \pm 0.49)$ with a rms scatter of 0.34 dex. $\gamma_{\rm{tot}}$ underestimates $\kappa_{\rm{tot}}$ by $\ge 2\times$ for four of 21 (19\%) of our fields.  If $\gamma_{\rm{tot}} \ge$ 0.01, then the $\kappa_{\rm{tot}}$ inferred directly from $\gamma_{\rm{tot}}$ would be underestimated by $\ge$ 0.01, leading to a $\ge$ 1\% bias in $H_0$.  The offset from our best fit $\kappa_{\rm{tot}}$ vs. $\gamma_{\rm{tot}}$ relation is not well-correlated with lens sightline galaxy number density, suggesting that it is difficult to simply correct $\gamma$ by galaxy density to get $\kappa$.

\end{itemize}

In summary, groups contribute significantly to the convergence in 15 of the 21 fields here.  These groups can be in the lens environment and/or along the line-of-sight.  For four fields, the lens group alone is significant. For 10 fields without a significant lens group, a line-of-sight group alone is significant.  For one field, several individually insignificant groups add to contribute a significant total convergence. 

Groups contribute significantly to the shear in 18 of the 21 fields analyzed.  For seven fields, the lens group alone is significant.  For nine fields without a significant lens group, a line-of-sight group alone is significant.  For four fields, several individually insignificant groups add to contribute a significant total shear.  For two fields, the total shear is not significant although an individual group's shear is because the shears of multiple groups cancel, as shears add as tensors.  These shears due to groups will affect overall shear measurements comparably or more than the 3$\sigma$ uncertainties in measured image positions.

To use galaxy-scale gravitationally lensed systems to determine cosmological parameters to the $\lesssim$ 1\% level--an LSST goal--it is critical to characterize and/or correct for both lens environments and line-of-sight structures.  For example, if the nine time delay lens systems in our sample are representative of those used in ensemble $H_0$ calculations, $H_0$ is overestimated by 11$^{+3}_{-2}$\% on average if both line-of-sight and the local lens environment are ignored. If only the line-of-sight contribution is ignored, then the $H_0$ bias is 7$^{+3}_{-2}$\%. Future surveys including volume-limited lens samples will reveal how general these results are.

\acknowledgments

We thank the anonymous referee for helpful comments.  M.L.W. and A.I.Z. acknowledge support from NSF grant AST-1211874.  M.L.W. also thanks the Technology and Research Initiative Fund (TRIF) Imaging Fellowship program for its support.  C.R.K. acknowledges support from NSF grant AST-1211385.  K.C.W. is supported by an EACOA Fellowship awarded by the East Asia Core Observatories Association, which consists of the Academia Sinica Institute of Astronomy and Astrophysics, the National Astronomical Observatory of Japan, the National Astronomical Observatories of the Chinese Academy of Sciences, and the Korea Astronomy and Space Science Institute.

{\it Facilities:} \facility{Mayall (Mosaic-I)}, \facility{Blanco (Mosaic-II)}, \facility{Magellan-1 (LDSS-2, LDSS-3)}, \facility{Magellan-2 (IMACS)}, \facility{MMT (Hectospec)}.

%\bibliography{grpgxysbibtex}{}
%\bibliographystyle{apj}

\clearpage

\LongTables
\begin{deluxetable*}{llrllrllr}
\tablecaption{Lensing Properties of Groups \label{table:totlensmax}}
\tablehead{
\colhead{Field} & \colhead{Group\tablenotemark{a}} & \colhead{$N_{m}$\tablenotemark{b}} & \colhead{$z_{\rm{grp}}$\tablenotemark{b}}& \colhead{$\sigma_{\rm{grp}}$\tablenotemark{b}} & \colhead{$R$\tablenotemark{c}}& \colhead{$\kappa_{\rm eff}$} & \colhead{$\gamma_{\rm eff}$} & \colhead{$\theta_{\gamma}$\tablenotemark{d}}\\
\colhead{} & \colhead{} & \colhead{} &  \colhead{} & \colhead{[km s$^{-1}$]} & \colhead{[$^{\prime}$]} & \colhead{} & \colhead{} & \colhead{[deg]} \\
}
\startdata

   b0712 &   0 &     5 &   0.0784 $^{+  0.0002}_{-  0.0003}$ &    140 $^{+    30}_{-   110}$ &   16.0 &   -                          &   -                         &   177 \\ 
         &   1 &     3 &   0.2635                       &    130                    &   12.6 &   -                         &   -                         &   104 \\ 
         &   2 &     3 &   0.3483                       &    110                    &    6.8 &   -                         &   -                         &   102 \\ 
         &   3 &    34 &   0.3723 $^{+  0.0007}_{-  0.0007}$ &    870 $^{+    70}_{-    70}$ &    7.7 &   0.002 $^{+  0.002}_{-  0.001}$ &   0.016 $^{+  0.005}_{-  0.004}$ &   150 \\ 
         &   4 &     4 &   0.3724 $^{+  0.0003}_{-  0.0002}$ &    170 $^{+    60}_{-   130}$ &   15.1 &   -                         &   -                         &    32 \\ 
         &   5 &     6 &   0.4146 $^{+  0.0003}_{-  0.0003}$ &    180 $^{+    40}_{-    60}$ &   14.1 &   -                         &   -                         &  -110 \\ 
         &   6 &     4 &   0.4360 $^{+  0.0004}_{-  0.0006}$ &    180 $^{+    30}_{-   150}$ &    2.8 &   -                         &   0.001 $^{+  0.001}_{-  0.001}$ &   -33 \\ 
         &   7 &     5 &   0.4938 $^{+  0.0007}_{-  0.0006}$ &    270 $^{+    50}_{-   230}$ &    4.8 &   -                         &   0.001 $^{+  0.001}_{-  0.001}$ &  -177 \\
\enddata
 
\begin{minipage}{12cm}
\tablenotetext{*}{Lens group.}
\tablenotetext{a}{Numbering as in \citet{wilson16}.}
\tablenotetext{b}{From \citet{wilson16}.  Uncertainties in group properties other than number of member galaxies are provided for groups with at least four members.}
\tablenotetext{c}{Projected distance between the group projected spatial centroid and the lens from \citet{wilson16}.}
\tablenotetext{d}{Angle measured north through east between the group projected spatial centroid and the lens.}
\tablenotetext{e}{Supergroup.}
\tablenotetext{f}{Main substructures of the supergroup.}
\end{minipage}

\end{deluxetable*}

\clearpage

\appendix

\section{Lens System References}

Here we present the references for the discoveries and lens and source redshifts for the systems in our sample.

\begin{deluxetable*}{lllll}
\tablecaption{Lens System References \label{table:lensrefs}}
\tablehead{
\colhead{Field} & \colhead{Official lens name} & \colhead{Lens discovery} & \colhead{$z_{L}$ reference} & \colhead{$z_{S}$ reference} 
}
\startdata
    b0712 & CLASS B0712+472 & \citet{jackson98} & \citet{fassnacht98} & \citet{fassnacht98} \\
    b1152 & CLASS B1152+200 & \citet{myers99} & \citet{momcheva15} & \citet{myers99} \\
    b1422 & JVAS B1422+231 & \citet{patnaik92} & \citet{kundic97} & \citet{momcheva15} \\
    b1600 & CLASS B1600+434 & \citet{jackson95} & \citet{fassnacht98} & \citet{fassnacht98} \\
    b2114 & CLASS B2114+022 & \citet{king99} & \citet{momcheva15} & \citet{augusto01} \\
  bri0952 & BRI 0952-0115 & \citet{mcmahon92} & \citet{eigenbrod07} & \citet{momcheva15} \\
  fbq0951 & FBQS J0951+2635 & \citet{gregg96} & \citet{eigenbrod07} & \citet{momcheva15} \\
    h1413 & H 1413+117 & \citet{magain88} & \citet{kneib98photz} & \citet{momcheva15} \\
\enddata

\begin{minipage}{12cm}
\tablenotetext{a}{https://www.cfa.harvard.edu/castles/}
\end{minipage}

\end{deluxetable*}

\section{Bootstrap Means Comparison}
\label{appendix:bootmean}

We would like to compare the mean values of several distributions in this work that are obviously non-Gaussian.  We therefore perform the following bootstrap method-based test.

Consider two distributions, $A$ and $B$, where the number of samples in $A$ ($N_A$) is larger than the number of samples in $B$ ($N_B$).  We randomly select $N_B$ values from distribution $A$ for 1000 trials.  We then find the fraction of trials in which the resulting mean values are at least as extreme as distribution $B$'s measured mean.  For example, if $B$'s measured mean is greater than the measured mean of $A$, we thus calculate the fraction of trials that result in means at least as large.

\section{Truncation Radius Choice}
\label{appendix:truncationchoice}

What group halo mass profile we assume, i.e., where or if we truncate it, could affect our results.  There is certainly mass associated with groups outside 1$r_{\rm{vir}}$. \citet{urban14} detect the Perseus cluster's intracluster medium beyond the virial radius in five of the eight radial directions they probe.  \citet{bahe13} and \citet{zinger16} have found cluster hot gas halos to extend beyond $1r_{\rm{vir}}$ (to at least 5$r_{200}$ and to $\sim 2-3 r_{\rm{vir}}$, respectively).  \citet{bahe13}, in their simulations, identify galaxies that are bound to the group and mostly have passed within 1$r_{\rm{vir}}$ but are on orbits with apocenters out to $\sim 3 r_{200}$, which motivated us in part to accept group members out to $3 r_{\rm{vir}}$ \citep{wilson16}.

The splashback radius is another possible halo edge definition.  This radius is where there is a steepening in the halo density profile due to a pileup of the apocenters of the matter that has been most recently accreted.  \citet{adhikari14}, using an analytical toy model, predict this radius to be $\sim$ 0.8-1$r_{200m}$ but found that it depends on accretion rate and redshift.  \citet{more15}, using simulations, also find that this radius depends on halo accretion rate but is typically around 0.8-1.5$r_{200m}$.  In another refinement, \citet{mansfield16} develop an algorithm to identify splashback shells that allows for asphericity.  They find slightly larger radii for halos with higher mass accretion rates and an overall splashback radius range of $\sim$ 1-1.6$r_{200m}$.  Observations suggest slightly smaller than expected splashback radii.  For example, \citet{more16} calculate splashback radii of $\sim$0.6-1$r_{200m}$ in their study of galaxy number densities around SDSS clusters.

It is unclear how well an NFW profile describes the outskirts.  NFW profiles were developed from virialized halos in simulations \citep{navarro97}; our groups might not all be relaxed (see Section \ref{sec:formalism}).  Outer cluster halos in simulations and some observations often have complex structures and are not always described well by NFW profiles.  For example, in their X-ray study of the Perseus cluster, \citet{urban14} find azimuthal variations and departures from expectations from numerical simulations; at $r > 0.4 r_{200}$, the average density profile's power law index is significantly smaller than predicted as well as there being differences in the entropy and pressure profiles.  They attribute these differences to gas being clumpy instead of smoothly distributed as models approximate.  However, the weak lensing analysis of \citet{umetsu11} find that, for the four of their five clusters not undergoing mergers, an untruncated NFW profile is consistent with their lensing measurements over their fields, which extend out to 1.2-1.7$r_{200}$.  \citet{umetsu16}, also using weak lensing, find that NFW profiles adequately describe the profiles of 16 Cluster Lensing and Supernova survey with Hubble clusters out to $\sim$ 2.5$h^{-1}$ Mpc ($\sim$ 1.5$r_{200m}$), outside of which the profile flattens.  In addition, the above simulations and observations mostly focus on cluster-sized halos, not the groups that make up most of our catalog.

As there is some ambiguity as to how to describe the mass outside $1 r_{\rm{vir}}$ for our groups and to the best description of the halo edge, we conservatively choose a truncation radius of 1$r_{200m}$ for our main analyses.

\begin{figure*}
\includegraphics[clip=true, width=18cm]{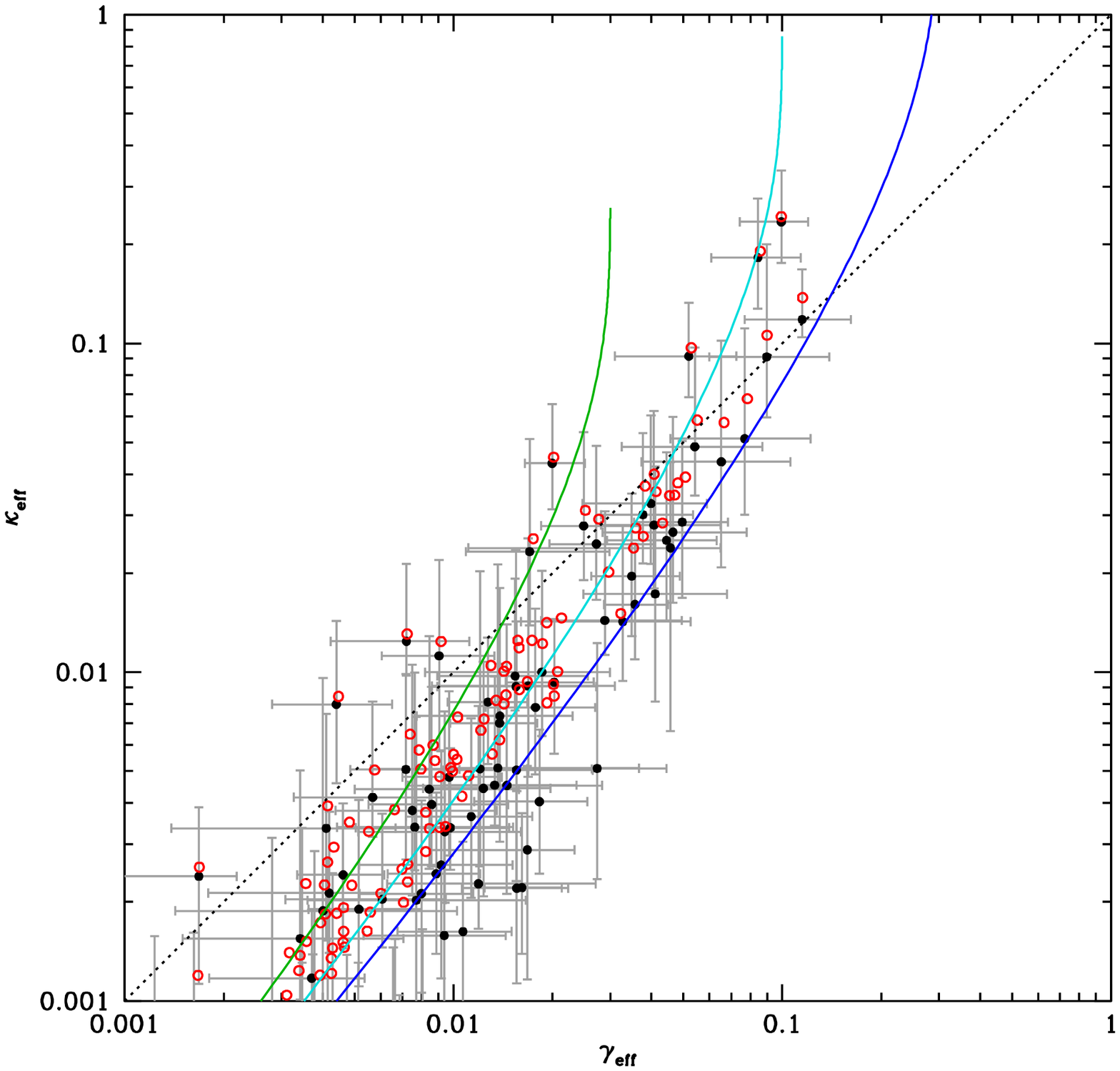}
\caption{Effective convergences and shears for all the groups in our catalog in fields other than h12531 (since this field has no source redshift estimate).  Black filled circles are values using NFW profiles truncated at $r_{200m}$, and red open circles are those using untruncated NFW profiles.  Gray error bars represent the 16th to 84th percentile ranges on the $\kappa_{\rm eff,truncated}$ and $\gamma_{\rm eff,truncated}$ values considering the uncertainties on the measured group properties.  The colored curves are the shears and convergences calculated with GRAVLENS \citep{keeton01} for untruncated NFW profiles at a range of projected separations from the lens for three values of $\kappa_s$ (0.03, 0.1, and 0.3 in green, cyan, and blue, respectively).  A dotted $\kappa = \gamma$ line is included for reference.  Truncating the halo results in smaller $\kappa_{\rm eff}$ values relative to $\gamma_{\rm eff}$.  However, at larger values of $\kappa_{\rm eff}, \gamma_{\rm eff}$, $\kappa_{\rm eff}$ grows quicker than $\gamma_{\rm eff}$ as separations from the lens decrease, so the relation converges to $\kappa = \gamma$.  Some of the scatter for a given truncation choice, especially at large values, is due to the range of $\kappa_s$ for our groups.
}
\label{fig:kappavsgammaefftruncvsnot}
\end{figure*}

We recalculate $\kappa_{\rm eff}$ and $\gamma_{\rm eff}$ using untruncated NFW profiles to see how this choice affects our results (see Figure \ref{fig:kappavsgammaefftruncvsnot}).

For either truncation choice, $\kappa_{\rm eff}$ is smaller than $\gamma_{\rm eff}$ at low values, although in general groups with higher $\kappa_{\rm eff}$ also have higher $\gamma_{\rm eff}$.  Convergence is a local quantity and depends on the projected mass surface density at the projected separation of the perturber's center from the main lens.  Shear, however, is a large-scale quantity that depends on the total projected enclosed mass as well as projected distance from the lens.  For a group projected much farther from the lens than the truncation radius, the mass surface density of the group halo projected directly in front of the lens will be quite small, so $\kappa_{\rm eff}$ will be very small.  $\gamma_{\rm eff}$ will also be small.  However, it will not be as small as $\kappa_{\rm eff}$, because the total projected enclosed mass does not get smaller with increasing distance from the group center like mass surface density does.  In the extreme case of a point mass projected at some distance from the lens, the convergence would be zero but the shear would be nonzero.

At high $\kappa_{\rm eff}$ and $\gamma_{\rm eff}$, $\kappa_{\rm eff}$ versus $\gamma_{\rm eff}$ is closer to a $\kappa = \gamma$ relation.  In the inner part of NFW halos, $\kappa_{\rm eff}$ diverges logarithmically while $\gamma_{\rm eff}$ converges to a constant.  So, as a halo is projected increasingly closer to the lens, the convergence grows more rapidly than the shear.  Thus, the convergence and shear can approach and pass the $\kappa = \gamma$ relation at very small projected distances from the lens.  However, it is unlikely for a group to be projected exactly in front of the lens, so few of our groups lie above the $\kappa = \gamma$ line.

Generally, the $\kappa_{\rm eff}$ for our groups using NFW halos truncated at $r_{200m}$ agree with those calculated using untruncated NFW profiles at $\kappa_{\rm eff,untruncated} \gtrsim 0.002$, and with $\gamma_{\rm eff,untruncated}$ values for the whole measured range, within the uncertainties in the values calculated using the truncated halos due to uncertainties in the measured group properties (see Section \ref{sec:formalism} for a description of our uncertainty estimation).  So which, if any, truncation radius we choose should not affect our results much more than the uncertainties in the group properties, although those uncertainties are random and these due to the truncation radius are systematic.  Those groups with $\kappa_{\rm eff,truncated} \ge 0.01$ and $\gamma_{\rm eff,truncated} \ge 0.01$ underestimate the untruncated values by a median fractional error of 24\% and 3\%, respectively; those with $0.001 \le \kappa_{\rm eff,truncated} < 0.01$ and $0.001 \le \gamma_{\rm eff,truncated} < 0.01$ have a median fractional error of 59\% and 22\%, respectively.

We repeat the main tests in our analysis, as well.  The quantitative results agree with each other within the uncertainties and do not affect our overall qualitative conclusions except in the following cases.

A larger fraction, 10 of 21 rather than 5 of 21, of our fields have $\kappa_{\rm{tot}}/\gamma_{\rm eff,los} \ge$ 2.  This still supports our result that the LOS group with the largest shear is a better predictor of the $\kappa_{\rm{tot}}$ than the lens group, but it is not as good a predictor as when truncated halos are used.

In the bootstrap realizations of the group catalog, slightly different fractions of the sample are found to have large differences between $\kappa$ and $\gamma$.  With untruncated halos, 24\% have $\kappa_{\rm{tot}}/\gamma_{\rm{tot}} \ge$ 2,  and 0.8\% have $\kappa_{\rm{tot}}/\gamma_{\rm{tot}} \ge$ 10. 72\% $\kappa_{\rm{tot}}/\gamma_{\rm eff,lens} \ge$ 2, and 19\% have $\kappa_{\rm{tot}}/\gamma_{\rm eff,lens} \ge$ 10.  40\% have $\kappa_{\rm{tot}}/\gamma_{\rm eff,los} \ge$ 2, and 4\% have $\kappa_{\rm{tot}}/\gamma_{\rm eff,los} \ge$ 10.  Our qualitative results that $\kappa$ is not a close predictor of $\gamma$ is unaffected, however.

Some of the difference between the values calculated using truncated versus untruncated halos might be reduced by including a void correction like that applied in \citet{mccully17}.  We do not investigate this issue quantitatively.  
  Since the choice makes little qualitative difference to our main results even without the void correction, it should not significantly impact our results.

\end{document}